\documentclass[12pt, preprint]{emulateapj}

\shortauthors{Shi et al.}
\begin{document}

\title{Role of Galaxy Mergers in Cosmic Star Formation History}

\author{Yong Shi\altaffilmark{1}, George Rieke\altaffilmark{1}, Jennifer Lotz\altaffilmark{2, 3}, Pablo G. Perez-Gonzalez\altaffilmark{4,5} }

\altaffiltext{1}{Steward Observatory, University of Arizona, 933 N Cherry Ave, Tucson, AZ 85721, USA}
\altaffiltext{2}{National Optical Astronomy Observatory, 950 N. Cherry Avenue, Tucson, AZ 85719, USA}
\altaffiltext{3}{Leo Goldberg Fellow}
\altaffiltext{4}{Departamento de Astrof\'{\i}sica, Facultad de CC. F\'{\i}sicas, Universidad Complutense de Madrid, E-28040 Madrid, Spain}
\altaffiltext{5}{Associate Astronomer at Steward Observatory, The University
of Arizona}

\begin{abstract}

We present  a morphology study  of intermediate-redshift (0.2$<z<$1.2)
luminous infrared  galaxies (LIRGs) and general field  galaxies in the
GOODS fields  using a  revised asymmetry measurement  method optimized
for deep fields.   By taking careful account of  the importance of the
underlying sky-background  structures, our new method  does not suffer
from systematic  bias and offers small  uncertainties.  By redshifting
local  LIRGs  and  low-redshift  GOODS galaxies  to  different  higher
redshifts, we  have found that  the redshift dependence of  the galaxy
asymmetry  due to  surface-brightness  dimming is  a  function of  the
asymmetry itself, with larger corrections for more asymmetric objects.
By applying redshift-, IR-luminosity- and optical-brightness-dependent
asymmetry corrections, we  have found that intermediate-redshift LIRGs
generally  show  highly  asymmetric  morphologies, with  implied  merger
fractions $\sim$ 50\% up to  z=1.2, although they are slightly more
symmetric than  local LIRGs.  For  general field galaxies, we  find an
almost  constant  relatively  high  merger  fraction  (20-30\%).   The
$B$-band LFs of  galaxy mergers are derived at  different redshifts up
to z=1.2 and  confirm the weak evolution of  the merger fraction after
breaking  the   luminosity-density  degeneracy.   The   IR  luminosity
functions  (LFs) of  galaxy  mergers are  also  derived, indicating  a
larger merger  fraction at higher  IR luminosity. The integral  of the
merger IR LFs indicates a  dramatic evolution of the merger-induced IR
energy  density [(1+z)$^{{\sim}(5{\textendash}6)}$],  and  that galaxy
mergers start to dominate   the cosmic IR energy density at z
$>$ $\sim$1.

\end{abstract}                                                    

\section{Introduction} 

Our  understanding  of galaxy  formation  and  evolution has  advanced
dramatically in the past decade, with well determined cosmic evolution
of the comoving star formation density, galaxy stellar mass and galaxy
metallicity content \citep[see][and references therein]{Cowie08}. What
mechanism drives this evolution?  Galaxy interactions and mergers play
a  key  role  in  the   {\it  theory}  of  galaxy  evolution,  through
transforming the galaxy morphologies, inducing violent starbursts, and
feeding the  central massive black  holes \citep{Mihos96, Springel05}.
However, it is  still unclear from observations to  what extent galaxy
mergers actually play the roles predicted for them.

Galaxy mergers  can be found from  observations either morphologically
or kinematically.  Morphologically-identified mergers include galaxies
with tidal  tails, wisps  and/or multiple nuclei.   They can  be found
through  either visual  classifications or  quantitative measurements.
While visual classification is the testbed to develop the quantitative
methods,  it is  time-consuming to  classify the  enormous  numbers of
galaxies found in deep surveys.  Moreover, distant galaxies can change
their  apparent morphologies due  to surface-brightness  (SB) dimming.
Such  an   effect  can   be  corrected  better   through  quantitative
measurements.   Several quantitative  morphology techniques  have been
developed,   CAS  \citep{Abraham94, Abraham96,   Conselice00,   Conselice03}  and
Gini-$M_{20}$  \citep{Abraham03, Lotz04}.   They  classify mergers  by
identifying galaxies  in a  pre-defined parameter space.  Although the
interpretation of  the parameter space is  somewhat uncertain, studies
based on the same quantitative definition can be compared and are less
subjective  than  visual  classification.   Galaxy  mergers  can  also
be identified  through kinematical  pairs  of galaxies  or galaxies  with
complicated  internal  velocity  fields  (i.e.  neither  pressure  nor
rotationally  supported).   Identifying  them requires  time-consuming
spectroscopic observations,  which have  been conducted for only  a limited
number of objects \citep{deRavel08, Neichel08}.

Different  approaches  to   merger  identification  are  sensitive  to
different stages of the galaxy merging process.  Pair-identification
algorithms   find  separated  interacting   galaxy  pairs   while  the
morphologically-based algorithms usually  identify galaxies during the
first  pass  and final  coalescence  where  the  galaxy morphology  is
significantly  disturbed  by  gravitational  torques  \citep{Lotz08b}.
Galaxy mergers with complicated interval velocity fields seem to span
a   longer  timescale  of   the  merging   process  \citep{Neichel08}.
Nevertheless,  no  method  can   identify all  mergers. In addition,
a fraction  of galaxies  identified as  mergers  by any  method are  not
necessarily  major  mergers.  For  example,  minor  mergers of  gas-rich
galaxies can also have highly disturbed morphologies (Lotz et al. 2009
in preparation).

Morphology  studies of  high-redshift  galaxies show  that the  Hubble
sequence is in  place by $z\sim$1 and that  high-redshift galaxies are
associated     more      frequently     with     peculiar     features
\citep[e.g.][]{Brinchmann98}.  Some studies have found that the galaxy
merger  fraction  shows strong  redshift  evolution, characterized  by
$(1+z)^{m}$  with   $m$  $>$  2   \citep[e.g.][]{LeFevre00,  Patton02,
  Conselice03,  Cassata05, deRavel08}, implying  that the  cosmic star
formation history (SFH)  is at least partly driven  by galaxy mergers.
Other studies, however, found a slow evolution \citep[e.g.][]{Bundy04,
  Lin08,  deRavel08}  or   even  no  evolution  \citep{Lotz08a}.   The
discrepancies  among different  studies  may be  caused by  low-number
statistics,  morphological  K-corrections  \citep[e.g.][]{Papovich03},
the method to identify mergers \citep{Lotz08b} and the physical properties
(e.g.    gas  fractions)   of   merger  progenitors   \citep{Lotz08b}.
Nevertheless,  almost all  studies  have found  relatively low  merger
fractions ($<$20\%) at $z$ $<$ 1.

A  more direct  way to  constrain the  role of  galaxy mergers  in the
cosmic  SFH  is  to  investigate  morphologies  of  luminous  infrared
galaxies (LIRGs, $L_{\rm IR}$  $>$ $10^{11}$ L$_{\odot}$).  LIRGs show
strong redshift evolution  and start to dominate the  cosmic IR energy
density at $z>$ 0.7 \citep{LeFloch05, Perez-Gonzalez05}.  In the local
universe,  LIRGs are  mainly ($\sim$50\%)  triggered by  major mergers
\citep{Sanders96}.       However,      morphology      studies      of
intermediate-redshift    LIRGs   indicate    low    merger   fractions
($\sim$10-30\%)    \citep{Zheng04,    Bell05,    Bridge07,    Lotz08a,
  Melbourne08}.  These findings have led to claims that the cosmic SFH
is driven  by some less violent  mechanisms, such as  accretion or gas
consumption \citep{Noeske07, Daddi08}.

\begin{figure}
\epsscale{1.}
\plotone{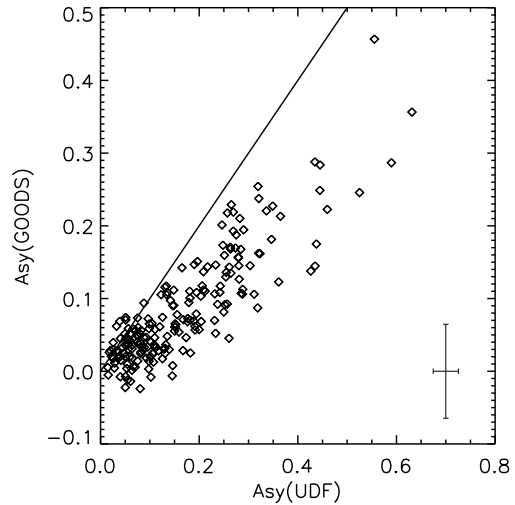}
\caption{\label{COMP_UDF_GOODS} Comparison of galaxy morphologies using
the GOODS $z$-band images and the UDF $z$-band images for about 250 UDF objects
with relatively high S/N and large size. The median error bar is shown in bottom-right corner. The solid line
is for Asy(GOODS)=Asy(UDF).}
\end{figure}

In \citet{Shi06},  we measured the  galaxy asymmetry for LIRGs  in the
Ultra Deep  Field (UDF),  which provides limiting  SB for  galaxies at
$z$=1 comparable  to that often  obtained for local ones  ($\mu_{B}$ =
$\sim$25.3  mag  arcsec$^{-2}$  at  10$\sigma$  in  the  AB  magnitude
system).  The merger fraction obtained  in this study is several times
higher than others, 40$\pm$24\% and 26$\pm$10 \% for LIRGs and general
field galaxies ($M_{B}$ $<$  -19.25) at $z\sim$0.7, respectively.  The
comparison   (see   Figure~\ref{COMP_UDF_GOODS})   between   asymmetry
measurements using the GOODS images  and ones using the UDF images for
the  same galaxies  indicates  substantial deficits  in the  indicated
asymmetry in  the shallower GOODS images.  These  two results motivate
us to  suspect that other  studies based on relatively  shallow imaging
(most   of  which   are  shallower   than  GOODS)   may  underestimate
significantly   the  role   of   SB  dimming   in  galaxy   morphology
measurements.   If so,  the  role of  mergers  in triggering  luminous
episodes of star formation may be seriously underestimated.

To  test the  result  of \citet{Shi06}  with  much higher  statistical
significance and  better constrain the  role of galaxy mergers  in the
cosmic SFH, in this paper we carry out detailed asymmetry measurements
and corrections  for all galaxies  ($\sim$16,000) with $m_{z}$  $<$ 25
including $\sim$7,500 galaxies and $\sim$1000  LIRGs at $z$ $<$ 1.2 in
the  GOODS field.   In \S~\ref{Gal-Str-Mea},  we describe  our revised
asymmetry estimate method.  In \S~\ref{Gal-Sam}, data for a complete local LIRG
sample and  the GOODS ACS-, redshift- and  {\it Spitzer}- observations
are presented, as well as the concentration and asymmetry measurements
for both  local LIRGs and  GOODS galaxies.  \S~\ref{RESULT}  shows the
evolution    of  the  observed     asymmetries    of    GOODS    galaxies
(\S~\ref{EVO-OBS-MOR})  and the  results after  asymmetry corrections,
including  the redshift  evolution  of the  merger  fraction in  LIRGs
(\S~\ref{Mor-Evo-LIRG}),  the  evolution  of  the merger  fraction  in
general  field galaxies  (\S~\ref{MOR-EVO-GAL}) and  the  infrared and
rest-frame   $B$-band   luminosity   functions   of   galaxy   mergers
(\S~\ref{IRLF-MERGERS}). In  \S~\ref{Discussion}, we discuss  the role
of  mergers in  galaxy star  formation activity  since  $z\sim$1.  The
conclusions  are in \S~\ref{conclusions}.   The technical  details are
given in  \S~\ref{New_Noise} for our  revised asymmetry method  and in
\S~\ref{Red-Dep-Gal-Asy} for asymmetry deficits of redshifted local
LIRGs   and    low-z   (z=0.2-0.4)   GOODS   galaxies    due   to   SB
dimming. Throughout the paper, we adopt a cosmology with $H_{0}$=70 km
s$^{-1}$ Mpc$^{-1}$,  $\Omega_{\rm m}$=0.3 and $\Omega_{\Lambda}$=0.7.
All magnitudes are defined in the AB magnitude system.

\section{Galaxy Structure Measurement}\label{Gal-Str-Mea}

\subsection{Galaxy Size And Concentration}

Quantitative   galaxy   morphologies   are  always   measured   within
well-defined  apertures in  order to  probe the same  physical  regions of
galaxies; these apertures  are optimized to include as  much galaxy
flux as possible while minimizing the effect of noise plus SB dimming.
When comparing  low- to high-redshift galaxy  morphologies, SB dimming
(SB $\propto$ $(1+z)^{4}$) can  artificially decrease the galaxy size.
To avoid such  an effect, the galaxy size can  be measured through the
dimensionless parameter:
\begin{equation}
\eta(r) =I(r)/{\langle}I(r){\rangle}
\end{equation}
where   $I(r)$  is   the  surface   brightness  at   radius   $r$  and
$\langle$$I(r)$$\rangle$ is the  mean surface brightness within radius
$r$ \citep{Petrosian76}.  In the standard approach \citep{Conselice00,
  Conselice03},  the  Petrosian  radius  $R_{\rm  p}$  is  defined  at
$\eta(r)$=0.2  and  the  galaxy  aperture  radius  is  defined  to  be
1.5$R_{\rm p}$.  The surface brightness $I(r)$ can  be measured within
either  circular  apertures  or  elliptical  apertures.  For  inclined
non-interacting  galaxies, the  mean galaxy  ellipticity  and position
angle  trace the  true galaxy  light  well.  The  $R_{\rm p}$  defined
within an  elliptical aperture is  generally larger than  that defined
within a circular  aperture and can be two  times larger for extremely
inclined systems.   However, we  have found that  interacting galaxies
usually   have  large   variations  in   their  position   angles  and
ellipticities.   The measured  galaxy  size for  them  depends on  the
adopted position  angle and ellipticity.  Because  of this uncertainty
for the  interacting systems, we adopt  circular apertures universally
for all galaxies in this paper.  For most local LIRG systems, we found
that  the difference  in $R_{\rm  p}$ between  circular  apertures and
elliptical  apertures  with mean  ellipticity  and  position angle  is
within 20\%.  Therefore, using  circular apertures does  not introduce
large errors.

The   concentration   parameter  measures   how   compact  the   light
distribution is.  It has been shown  to correlate  well with the
Hubble  type \citep{Kent85,  Bershady00}.  Early-type galaxies
generally have  more compact light distributions  than do late-type
ones. With the curve of growth measured within the galaxy aperture, the
concentration can be  defined as the ratio of the two radii  enclosing
two fixed fractions of the light.  Following  \citet{Kent85} and 
\citet{Bershady00},  we have  used the  following definition  in  this paper:
\begin{equation}
C=5\log{\frac{R_{80}}{R_{20}}},
\end{equation}
where $R_{80}$ and  $R_{20}$ are the radii that  enclose 80\% and 20\%
of the total light, respectively.

\subsection{Revised Asymmetry Parameter}\label{Revised_Asy}

%%Visually identified  galaxy mergers are
%%biased  to   high  SB  objects especially at  high redshift.  It  is  impossible   with  visual
%%classification  to do  any systematic  corrections for  missing  low-SB
%%asymmetric structures  at high redshift.
The asymmetry parameter ($A$) was first introduced by \citet{Abraham96}
and  subsequently developed  by  \citet{Conselice00, Conselice03}.   It
quantitatively describes  the level  of the galaxy asymmetry  and provides a
direct measurement  of the importance of  asymmetric sub-structures in
galaxy morphologies,  such as  merging companions  and tidal tails,
which  are  used  to  identify   mergers  in  visual
classification.   Accurate  measurements   of  the asymmetry  parameter are thus
critical to understand the role of galaxy mergers in galaxy evolution.
 The
galaxy asymmetry is defined as:
\begin{equation}
A_{\rm galaxy+noise} = \frac{{\sum}|I_{\rm galaxy+noise} - I_{\rm galaxy+noise}^{180^{\circ}}|}{{\sum}I_{\rm galaxy+noise}}
\end{equation}
where  $A_{\rm galaxy+noise}$  is  the asymmetry  of  the  galaxy
signal plus the  noise, $I_{\rm galaxy+noise}$ is the  image of galaxy
signal  plus  the noise  and  $I_{\rm galaxy+noise}^{180^{\circ}}$  is
$I_{\rm  galaxy+noise}$ with the image  rotated by  $180^{\circ}$  around a  rotation
center. The true pure galaxy asymmetry is then given by:
\begin{equation}
A_{\rm galaxy} = {\rm min}(A_{\rm galaxy+noise}) - A_{\rm noise}^{\rm corr}
\end{equation}
where ${\rm min}(A_{\rm galaxy+noise})$ is the minimum of the galaxy signal plus noise
asymmetry, and $A_{\rm noise}^{\rm corr}$ is the noise correction obtained by rotating
a background image around a center and normalizing by the galaxy brightness:
\begin{equation}
 A_{\rm noise}^{\rm corr} = \frac{{\sum}|B - B^{180^{\circ}}|}{{\sum}I_{\rm galaxy+noise}}
\end{equation}
where $B$ is the background image without any object. 

 The overall measurements of galaxy concentration and asymmetry are  composed of the following steps:\\
(1) Guess the initial center. \\
(2) Measure the $R_{p}$ and concentration. \\
(3) Search for the rotation center that gives the minimum of $A_{\rm galaxy+noise}$ within a 1.5$R_{p}$ aperture.\\
(4) Use the new rotation center to re-measure the $R_{p}$ and concentration.\\
(5) Use the above rotation  center and new $R_{p}$ to search for a new rotation
center that gives the minimum of $A_{\rm galaxy+noise}$ within a 1.5$R_{p}$ aperture.\\
(6) Correct for the noise $A_{\rm noise}^{\rm corr}$.

The galaxy centers can  be easily located for non-interacting galaxies
but  not for  interacting ones.  Here,  we adopt  the rotation  center
giving  the minimum of  $A_{\rm galaxy+noise}$  to measure  the galaxy
size as is done in steps  1-3. Note that the asymmetry rotation center
in step 5  is usually not very  different from that in step  3, as the
``walking around'' method invented by \citet{Conselice00} is generally
robust to  find the  minimum $A_{\rm galaxy+noise}$.   For discussions
about  the galaxy  size definition,  the algorithm  searching  for the
minimum   $A_{\rm   galaxy+noise}$,  and  the   dependence   of   $A_{\rm
  galaxy+noise}$   on  the  resolution   and  correlated   noise,  see
\citet{Conselice00, Conselice03}.

The  technical details about  our new  $A_{\rm noise}^{\rm  corr}$ are
given  in  \S~\ref{New_Noise}.   Here  we  present a  summary  of  the
procedure   to   determine   this    parameter.    A   set   of   1000
randomly-produced noise asymmetry  measurements is carried out by putting
circular regions in the background image around the target.  The value
of this distribution  at the 15\% probability low-end  tail is used as
$A_{\rm noise}^{\rm corr}$, the  median of true noise corrections. The
error  ($\sim$3$\sigma$) in  the  final measured  galaxy asymmetry  is
taken   to   be   two   times   the  standard   deviation   of   these
randomly-produced  noise asymmetries.  In  reality, some  galaxies are
always present in the field of targets. To account for this problem, a
success  rate   is  defined  for  the   one-thousand  circular  region
placements as  the fraction of  circular regions containing  no galaxy
signal  indicated  by  the  SExtractor segmentation  image.   Circular
regions  containing any  galaxy signal  are not  used.  More  sets of
one-thousand  placements  are generated  until  one  thousand  successful
measurements are  reached.  If the success  rate for one  set of 1000
placements  is lower  than  50\%,  the circular  region  size for  the
following set  is taken  to be  80\% of this  set.  Then  the measured
noise asymmetry is rescaled to that with the original size by assuming
that the noise asymmetry is proportional to the aperture area.

\section{Galaxy Sample}\label{Gal-Sam}
\subsection{Local LIRG Sample}\label{Gal-Sam-LIRG}

Local LIRGs and ULIRGs  are generally galaxies undergoing mergers with
morphological signatures  of tidal  tails, multiple nuclei  and highly
asymmetric   features   \citep{Sanders96}.    To   account   for   the
redshift-dependence of  galaxy merger morphologies due  to SB dimming,
we  measured  the morphologies  of  local  LIRGs  and of  local  LIRGs
redshifted   to   different  redshifts.    We   retrieved  {\it   HST}
ACS/WFC-F435W images of 88 local  LIRGs observed in Program ID - 10592
(PI - Aaron  Evans).  This local LIRG sample  is a complete sub-sample
of  the   IRAS  Revised   Bright  Galaxy  Sample   \citep[RBGS:  i.e.,
  $f_{60{\mu}m}$    $>$   5.24    Jy;][]{Sanders03}    above   $L_{\rm
  IR}=10^{11.4}$  L$_{\odot}$.  The  sample size  is large  enough for
statistically valid  comparisons.  The distance of  this sample covers
the range  from 35 to $\sim$350  Mpc.  The median distance  of 135 Mpc
corresponds to  a physical  resolution of 80  pc.  These  galaxies are
also bright enough  to be useful when redshifted  to higher redshifts.
These  local  LIRGs  are  currently  experiencing  merging  processes,
spanning a  wide range of merging stages  from well-separated galaxy
pairs to the final merging remnants. 

To obtain  quantitative morphology measurements,  foreground stars and
background galaxies  should be removed  from the images  first.  Given
the  importance  of  the  background  region  in  accurate  morphology
measurements  (see  \S~\ref{New_Noise}),  we  carried  out  a  careful
removal of contaminators, especially given that some of the LIRGs have
tens  of foreground  stars.   We used  SExtractor \citep{Bertin96}  to
obtain the segmentation map of  each image.  To make sure the extended
low-SB  emission was  included in  the object,  we first  rebinned the
image by  a factor of 4$\times$4  and then adjusted  the parameters in
SExtractor so  the extended emission was  detected.  This segmentation
map was then resampled to the original resolution by assuming that the
original   pixels  belonging   to  a   rebinned  pixel   conserve  the
segmentation value. We then cleaned the neighbourhood of the target by
replacing  pixels  of  non-target object pixels randomly  with  values  of
background pixels.   As  the   neighbourhood  is  dominated  by  the
background pixels ($>$ 95\%),  such replacement does not introduce any
significant correlation in the final cleaned neighbourhood background.
Sometimes  a target  was identified  as several  different  objects by
SExtractor.   We  visually inspected  each  image  to  make sure  such
substructures   were  still   included  in   the  target   during  the
neighbourhood  clean. Sometimes  foreground  stars lie  within the
target region and cannot be separated by SExtractor.  We used the IRAF
package IMEDIT to remove these contaminators further.

As  the local  LIRG sample  is a  flux-limited sample  instead  of a
volume  limited sample,  we have  applied weights  of $\frac{1}{V_{\rm
    max}}$/$\sum$$\frac{1}{V_{\rm max}}$ to  each object to obtain an equivalent
volume-limited  asymmetry distribution, where  $V_{\rm max}$
is  measured  at  the  redshift  where  the object  with  a  given  IR
luminosity has $f_{60{\mu}m}$ $=$ 5.24 Jy.

\subsection{Galaxies in the GOODS Field}
\subsubsection{{\it HST} Images and Morphology Measurements}

The GOODS  field consists of two sub-fields,  GOODS-North (GOODS-N) and
GOODS-South (GOODS-S),  imaged with {\it HST}/ACS in  four filters $B$
(F435W),  $V$   (F606W),  $i$  (F775W)  and   $z$  (F850LP)  \citep[M.
  Giavalisco and the  GOODS Team, 2008, in preparation]{Giavalisco04}.
The field centers (J2000.0) are 12h36m55s, 62$^{\circ}$14$'$15$''$ for
the  GOODS-N and  3h32m30s, -27$^{\circ}$48$'$20$''$  for  the GOODS-S.
The survey area is 320  arcmin$^{2}$ with $BViz$-band coverage and 365
arcmin$^{2}$  with  $Viz$-band  coverage for each field.   GOODS Version  2.0  provides
exposure times  of 7200, 5650, 8530  and 24760 secs in  GOODS-N and
7200, 5450, 7028 and 18232 secs  in  GOODS-S. The final image has a
pixel  scale of  0.03$''$/pixel. The  GOODS-S and  GOODS-N  fields are
divided into 18 and  17 sections, respectively. An individual image file
is released for each section.

Objects   were   detected   in   $z$-band   with the  SExtractor   package
\citep{Bertin96}  and photometry was  measured in  all four  bands.  We
produced the segmentation images  that define the galaxy pixels using
the  released parameter files  of SExtractor plus  weight map  images, and
final science images.  We used galaxy magnitudes defined by SExtractor
MAG$\_$AUTO.

We carried  out morphology measurements in the  $Viz$-bands for 16,708
objects  with  $m_{z}$  $<$  25,  SExtractor  CLASS$\_$STAR  $<$  0.9,
FLUX$\_$RADIUS1  $<$ 100,  IMAFLAGS$\_$ISO $<$  16 and  not  within 33
pixels of  the edge of each SECTION  field. SExtractor FLUX$\_$RADIUS1
is the radius in pixels enclosing 20\% of the flux and FLUX$\_$RADIUS1
$<$  100 excludes  9 artificial  objects that  are long  narrow bright
belts  across  the field.   IMAFLAGS$\_$ISO  $<$  16 excludes  objects
within 33  pixels of the  field edge.  The morphology  measurement was
first  carried out in $z$-band  starting  with  cutting  out science  and
segmentation images for  each object. As the galaxy  radius is defined
to  be 1.5$R_{p}$  and  the asymmetry  uncertainties  are measured  by
putting  circular regions  randomly in  background regions  around the
objects,   the    image   size   was    cutout   with   a    size   of
9$R_{p}{\times}$9$R_{p}$.    Then  the   $z$-band   concentration  and
asymmetry   were  measured   as  described   in  \S~\ref{Revised_Asy}.
Briefly, the residual  sky was first subtracted using  the mean values
of all pixels with  zero segmentation values.  Companions were defined
as  all pixels  with non-zero  segmentation  values not  equal to  the
target's  value. They  were removed  by replacing  their  pixel values
randomly with those of sky pixels. Note that the removed companions are
those well separated from targets and thus do not contribute to the target 
asymmetries. The initial center was given by the
astrometrical   position   of   the   object.   The   first   set   of
concentration/asymmetry measurements  gave new centers  and the second
measurements  were  carried out  to  give  $R_{p}$, concentration,  and
asymmetry  values.  The  asymmetry  uncertainty was  measured in  1000
circles  randomly  put  in   the  sky  region,  excluding  object  and
companions    (for    details,    see    the   last    paragraph    of
\S~\ref{New_Noise}).    To  evaluate   the  dependence   of the  asymmetry
uncertainties on the S/N, two types  of S/N are defined. The total S/N
is the total  signal divided by the total sky  noise for galaxy pixels
within the  1.5$R_{p}$ radii. Note  that sky and companion  pixels are
excluded.  Similar to \citet{Lotz08a},  $\langle$S/N$\rangle$ is defined to be the
arithmetic  (not  quadratic)  average  of  the S/N  of  each  galaxy
pixel. The asymmetry/concentration measurements failed for 212 objects (175 of them
are stars; the remaining  37 objects  either have extremely
low  SB  or are near  the edge, within 50  pixels).  No systematic
correction is made for these objects, as the SB limiting cut will
be applied in  the final morphology catalog and they  are a very small
fraction    (0.2\%)   of   the    total number of  objects.     The   $Vi$-band
concentration and asymmetry were then  measured for all  objects with successful
$z$-band concentration and asymmetry measurements.

\subsubsection{{\it Spitzer} MIPS and IRAC Data}

MIPS and IRAC data for the  GOODS-N and GOODS-S fields were drawn from
the {\it  Spitzer} GTO observations  of the larger area  covering each
field,     1.5$^{\circ}$$\times$0.5$^{\circ}$     for     MIPS     and
1.0$^{\circ}$$\times$0.5$^{\circ}$ for IRAC.   The reduction of the 24
$\mu$m  images  was carried  out  with  the  MIPS Data  Analysis  Tool
\citep{Gordon05}.  The  detailed data reduction,  object detection and
photometry measurement procedures  are given in \citet{Papovich04} and
\citet{Perez-Gonzalez05}.  The  final MIPS  24 $\mu$m catalog  is 50\%
complete  at   60  $\mu$Jy.  

\subsubsection{Redshifts And Morphology-Redshift Catalog}\label{RED_MOR_RED_CAT}

For the GOODS-S field,  the spectroscopic redshifts were obtained from
Version 3.0 of the FORS2 catalog \citep{Vanzella08} and Version 1.0 of
the  VIMOS  catalog  \citep{Popesso08}.   Only  solid  redshifts  with
redshift  quality $z_{q}$=$'$A$'$  were  used.  For  GOODS-N, we  used
spectroscopic  redshifts  with  $z_{q}$=3  or  4 from  the  Team  Keck
Treasury Redshift  Survey (TKRS)  catalog. For objects  without secure
spectroscopic     redshifts,     we     used    the     catalog     of
\citet{Perez-Gonzalez08},  who obtained  photo-z  based on  photometry
covering    from   UV    to   IRAC    bands.    The    photo-z   error
$\sigma$$_{z}$/(1+$z$) is  $<$ 0.2 for  95\% of the redshifts  and $<$
0.1 for  88\% of the  redshifts. The median  $\sigma$$_{z}$/(1+$z$) is
0.03.  The  photo-z error  is small enough  for our purpose,  i.e., to
determine  the rest-frame band  of the  galaxy image.   These redshift
catalogs were matched to the GOODS $z$-band morphology catalog using a
search radius of 0.5$''$. The redshift was assigned to the nearest one
if multiple objects were present  within a search radius. The redshift
completeness for $m_{z}$  $<$ 25 is 68\%; it  is mostly determined by
the  IRAC detections  for which  the  photo-z can  be calculated  in
\citet{Perez-Gonzalez08}.  We refer to the sample of objects with both
morphology  and  redshift   measurements  as  the  morphology-redshift
catalog.

To account for the  morphological K-correction, the redshift evolution
of the galaxy morphology was determined in the rest-frame $B$-band for
objects at $z$ $<$ 1.2. The  redshift range of $z$ $<$ 1.2 was divided
into five  redshift bins,  [0.2, 0.4], [0.4,  0.6], [0.6,  0.8], [0.8,
  1.0]  and [1.0,  1.2].  The  rest-frame $B$-band  morphologies  in a
given  redshift bin  were  defined  as the  morphologies  in the  band
nearest to the redshifted rest-frame $B$-band, i.e, the observed-frame
$V-$, $V-$, $i-$, $i-$ and $z$-band  for the redshift bins from low to
high.

\begin{figure}
\epsscale{1.}
\plotone{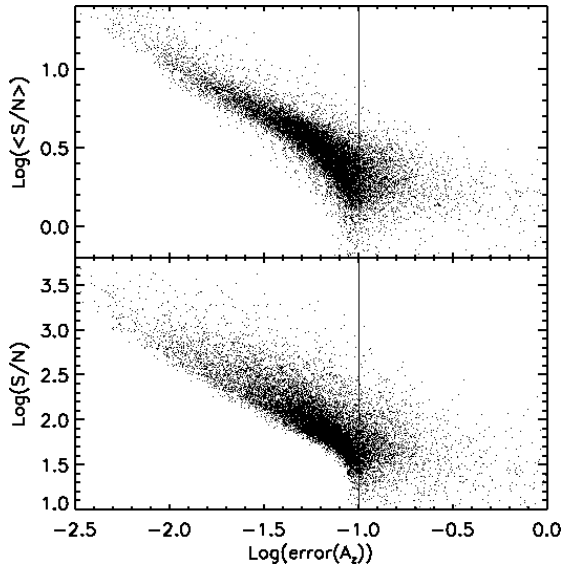}
\caption{\label{S_N_Asyerror} The S/N vs. asymmetry uncertainty (3$\sigma$), where the upper panel
shows the mean S/N and the lower panel is for the total S/N. The S/N is only 
correlated with the asymmetry uncertainty for galaxies with S/N $>$ 30 and $\langle$S/N$\rangle$ $>$ 3.   }
\end{figure}

To  obtain the  optical counterparts  of  the MIPS  sources, we  first
obtained the IRAC 8$\mu$m counterparts using a search radius of 1$''$.
Then for MIPS sources with IRAC counterparts, the optical counterparts
were defined as objects in  the morphology catalog within 1$''$ of the
IRAC  counterparts.   For  MIPS  sources  without  IRAC  counterparts,
optical  counterparts were  searched for  within 2.5$''$  of  the MIPS
source.  For multiple objects within  a search radius, the nearest one
was defined to be the optical counterpart. For MIPS sources with known
optical redshifts, the total  8-1000 $\mu$m IR luminosity was obtained
based  on the  observed 24  $\mu$m flux  density and  the star-forming
templates from \citet{Rieke09}.

\subsection{Reliable Morphology And Completeness Cut}\label{Complete_CUT}

\begin{figure}
\epsscale{1.0}
\plotone{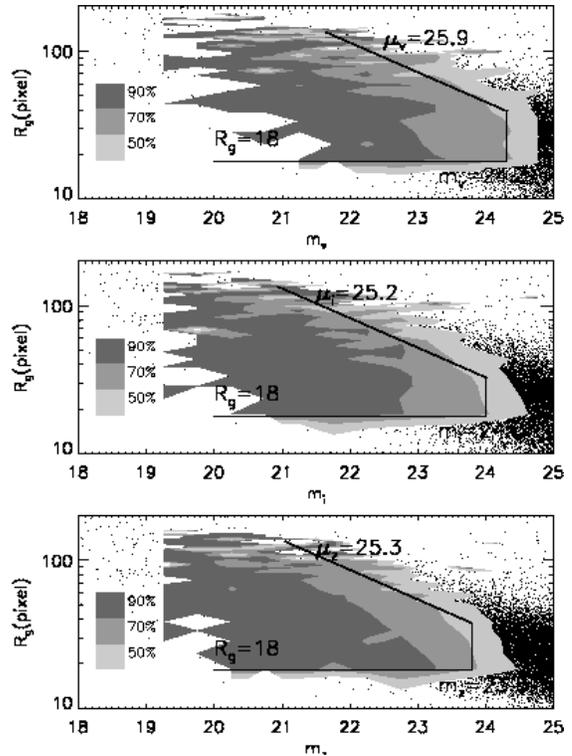}
\caption{\label{Completeness_MOR_RED}  The galaxy radius vs. magnitude in $v$-, $i$-, $z$-
bands for galaxies selected with $m_{z}$ $<$ 25. Areas with different grayscales correspond to different completeness
cuts of galaxies with reliable morphology and redshift measurements (see \S~\ref{Complete_CUT}). The solid lines
enclose the 70\% complete area where the apparent surface brightness, size and magnitude are labelled. }
\end{figure}

The redshift-morphology  catalog was  further limited to  objects with
reliable  asymmetry  measurements,   defined  as  those  with  $R_{\rm
  gal}$=1.5$R_{p}$ $>$  15 pixels (0.03$''$/pixel)  and error($A$) $<$
0.1 in a  given band. The accuracy of the  galaxy asymmetry depends on
the resolution and S/N.  For  a galaxy with small size, the resolution
is  not high  enough to  resolve galaxy  structures and  the asymmetry
becomes artificially small  \citep{Conselice00}. For galaxies with low
S/Ns, the  asymmetry uncertainty is large  due to the  noise.  For the
study in this paper, we  used an asymmetry uncertainty cut (error($A$)
$<$ 0.1 at approximately 3$\sigma$ level) instead of a S/N cut, as our
revised  asymmetry  method can  characterize  the  uncertainty due  to
complicated  background  structures.  Figure~\ref{S_N_Asyerror}  shows
the  advantage of  the asymmetry  uncertainty  cut over  the S/N  cut.
While both  S/N and $\langle$S/N$\rangle$ are  tightly correlated with
the asymmetry uncertainty at high S/N (e.g., $\langle$S/N$\rangle$ $>$
3),   the    scatter   of    the   correlation   becomes    large   at
$\langle$S/N$\rangle$    $<$    3.    A    high    S/N   cut    (e.g.,
$\langle$S/N$\rangle$ $>$ 3) will exclude many objects with relatively
small  asymmetry uncertainty  (error(A) $<$  0.1)  and a  low S/N  cut
(e.g., $\langle$S/N$\rangle$ $>$ 1)  will introduce objects with large
asymmetry  uncertainty (error(A)  $>$  0.1).  Our  error($A$) $<$  0.1
criterion    only    excludes    objects    with    large    asymmetry
uncertainties. Most  objects with  error($A$) $>$ 0.1  have relatively
inhomogeneous backgrounds.  All objects with error($A$) $>$ 0.3 are at
the edge  of the  field where the  exposure is relatively  shallow and
shows a large  gradient toward the edge.

We now  have a  redshift-morphology catalog with  redshift information
and reliable morphology measurements. The redshift evolution of galaxy
morphologies will be characterized in a subsample of this catalog that
is  complete to  a  certain value.   Figure~\ref{Completeness_MOR_RED}
shows  the   completeness  cut   at  50\%,  70\%   and  90\%   of  the
redshift-morphology catalog  in $R_{\rm gal}$ vs.   magnitude in three
bands, where the completeness is defined as the ratio of the number of
galaxies  with redshift  measurements and  secure morphologies  to the
total number of  galaxies in a $R_{\rm gal}$-magnitude  bin. The  absolute
values     at    different    redshift     bins    are     given    in
Table~\ref{Tab_Completeness}.    Here,   we   do   not   account   for
incompleteness of the {\it  HST}/ACS catalog, as its incompleteness at
$m_{z}$ $<$ 25 is negligible.

\begin{figure}
\epsscale{1.0}
\plotone{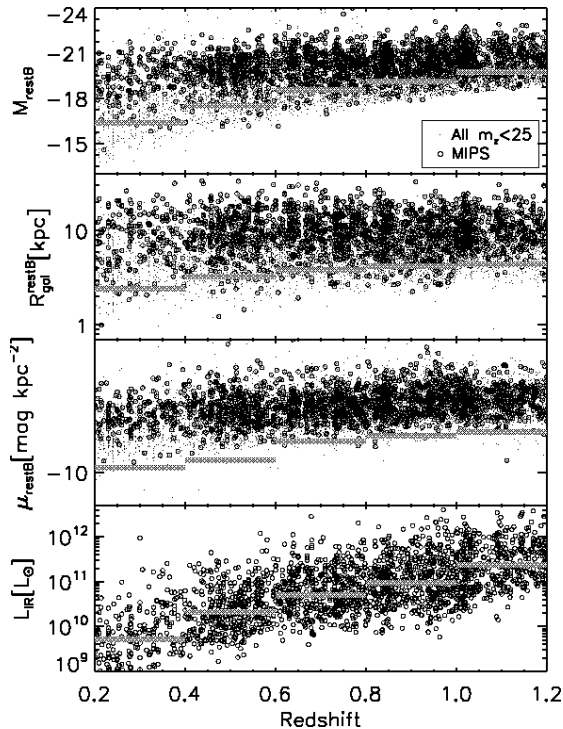}
\caption{\label{MB_LIR_red} The first three panels show  
the rest-frame $B$-band magnitude,  the galaxy radius and the rest-frame 
$B$-band surface brightness as functions of redshift for 
all galaxies with $m_{z}$ $<$ 25 (dots) and MIPS-detected galaxies (open circles).
The heavy solid lines correspond to the 70\% completeness cut of the redshift-morphology sample (see 
\S~\ref{Complete_CUT}). The last panel shows the total IR luminosity
of MIPS-detected galaxies as a function of redshift where the solid lines
correspond to the 50\% completeness cut of the 24$\mu$m detections.}
\end{figure}

Figure~\ref{MB_LIR_red}  shows  the  absolute  magnitude,  galaxy  size,
surface brightness and IR luminosity as a function of redshift for all
$m_{z}$ $<$  25 and MIPS-detected objects  with redshift measurements.
The absolute rest-frame $B$-band magnitude is measured through the version
4.1.4 KCORRECT code \citep{Blanton03, Blanton07} using four ACS bands.
For the first  three quantities, the 70\%-complete redshift-morphology
cut is drawn  in each redshift bin and the 50\%-complete  IR flux limit is drawn
for the IR luminosity.  As  shown in the figure, most objects with $m_{z}$
$<$ 25 and  $z$ $<$ 1.2 are included  in the final redshift-morphology
catalog.   This   is   because   the   incompleteness   of   the
redshift-morphology   catalog  is  mainly   caused  by   the  redshift
incompleteness instead of the criterion of secure morphologies.

\section{RESULT}\label{RESULT}

\subsection{Observed Asymmetry Distributions of GOODS Galaxies}\label{EVO-OBS-MOR}

\begin{figure*}
\epsscale{1.0}
\plotone{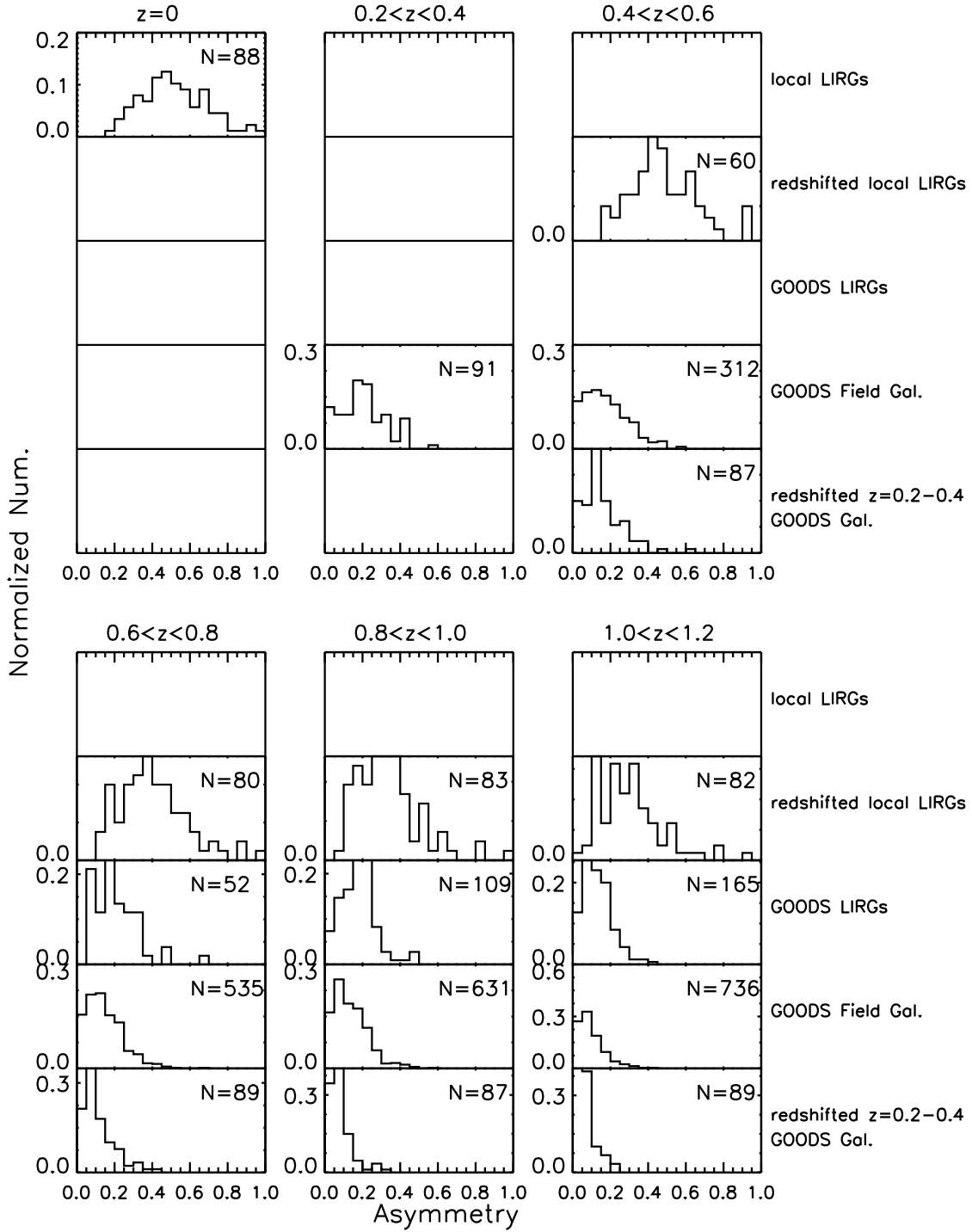}
\caption{\label{Asy_Dis_z}  Comparisons in the  asymmetry distribution
between   local  LIRGs   ($L_{\rm   IR}$  $>$   2.5$\times$10$^{11}$
$L_{\odot}$), redshifted local LIRGs,  GOODS LIRGs ($L_{\rm IR}$ $>$
2.5$\times$10$^{11}$   $L_{\odot}$),   GOODS   field  galaxies ($M_{B}$ $<$ -19.75)   and
redshifted low-z(z=0.2-0.4) GOODS galaxies within different redshift
bins.   Within  each redshift  bin,  the  asymmetry distribution  on
average becomes progressively smaller from the top to the bottom.}
\end{figure*}

\begin{figure*}
\epsscale{1.}
\plotone{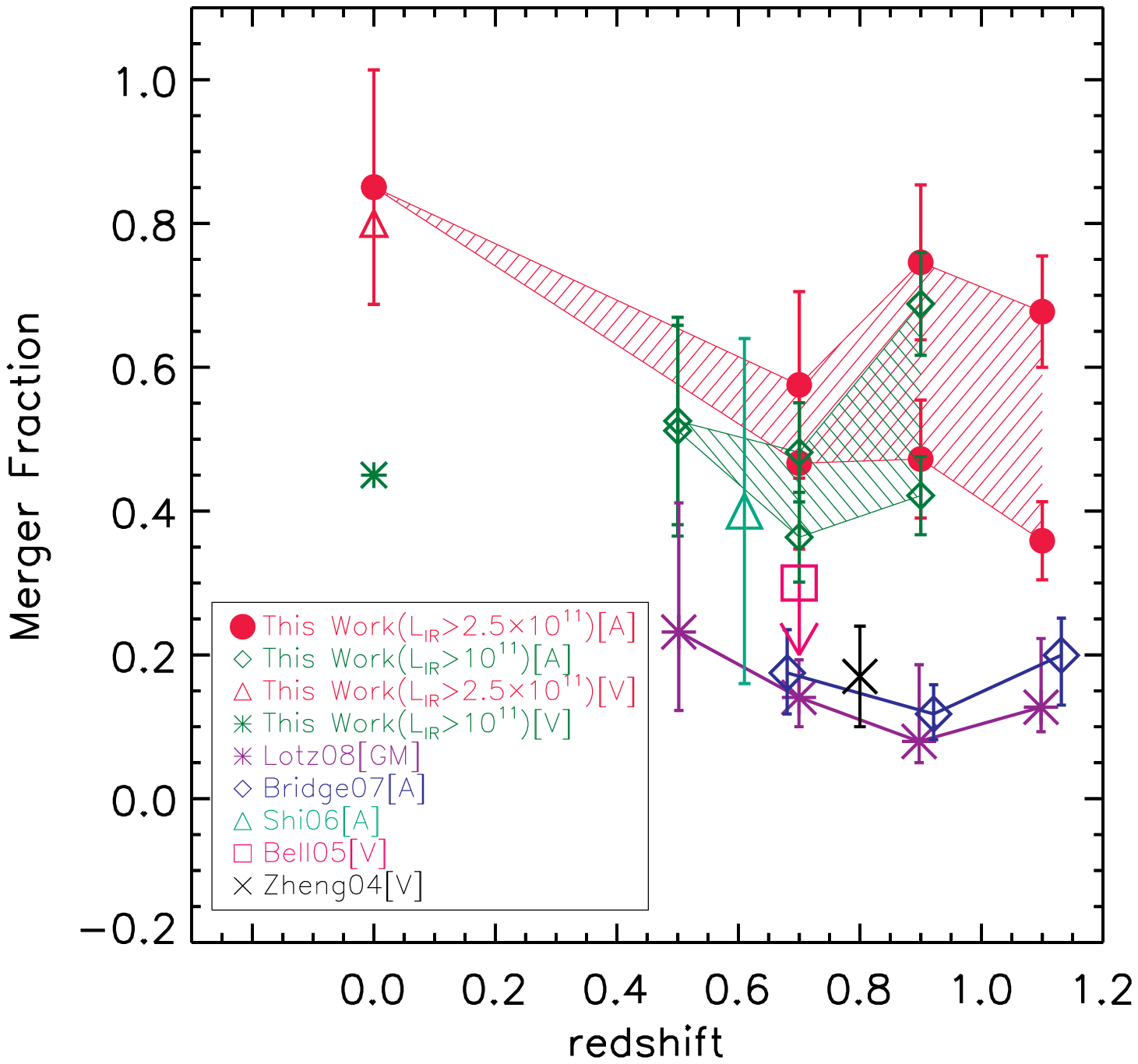}
\caption{\label{MergeFrac_Red}   Redshift  evolution  of   the  merger
fraction  in LIRGs,  where  upper- and lower-limits at  a given redshift correspond to  the result with
asymmetry corrections based on redshifted local LIRGs and redshifted
low-z GOODS  galaxies, respectively. Our  work is given with  two IR
luminosity cuts, i.e., $L_{IR}$ $>$ 2.5$\times$10$^{11}$ L$_{\odot}$
and $L_{IR}$  $>$ 10$^{11}L_{\odot}$.  All other works  are given at
$L_{IR}$ $>$  10$^{11}L_{\odot}$.  $''A''$: mergers  identified with
asymmetry;  $''$GM$''$: Gini-$M_{20}$  classified  mergers; $''V''$:
visually classified mergers.}
\end{figure*}

Two main results described in the Appendix are important for the study
of morphology  evolution.  First,  as shown in  \S~\ref{New_Noise}, we
found that the underlying structure of the sky background is important
in accurate asymmetry measurements  and we propose a revised asymmetry
measurement approach.  Secondly, by  redshifting local LIRGs and low-z
(z=0.2-0.4) GOODS galaxies to higher redshifts, asymmetry deficits due
to  SB dimming  for  these two  galaxy  populations are  derived as  a
function  of   redshift.   As  shown   in  \S~\ref{red-dep-asym-LIRG},
originally  more asymmetric  objects show  larger  asymmetry deficits.
Such dependence  indicates that to recover  the intrinsic distribution
of  a galaxy  population at  a given  redshift, a  low-z galaxy
population  with  intrinsically  similar  asymmetry should  be  used  to
construct  the   asymmetry  corrections.   Such a  low-z  galaxy
population can  be defined  through comparing the observed  asymmetries of
redshifted low-z galaxies to  those of the galaxy population at
a given  redshift, as the observed asymmetries still  on average correlate
with original asymmetries even given the larger asymmetry deficits for originally
more asymmetric objects (see Figure~\ref{COMP_UDF_GOODS}).

Figure~\ref{Asy_Dis_z}  shows the observed  asymmetry of  local LIRGs,
GOODS LIRGs  and GOODS field galaxies at  different redshifts compared
to the observed asymmetry of redshifted local LIRGs and redshifted low-z
(z=0.2-0.4) GOODS galaxies. The GOODS LIRGS are the objects at $L_{\rm
  IR}$  $>$ 2.5$\times$10$^{11}$ $L_{\odot}$  and satisfying  the 70\%
completeness   cut   of  the   redshift-morphology   catalog  at   the
corresponding  redshifts  (see  Figure~\ref{Completeness_MOR_RED}  and
Table~\ref{Tab_Completeness}).   The  GOODS  field galaxies  within  a
redshift interval  are all  galaxies satisfying the  70\% completeness
cut of the  highest redshift bin.  For the  redshifted local LIRGs and
low-z  GOODS galaxies,  the observed  asymmetry  becomes progressively
smaller at higher  redshift as caused by SB  dimming (see Appendix for
discussion).    

Figure~\ref{Asy_Dis_z}  shows a  trend of  the observed  asymmetry for
different  subsets of the  galaxy population  within a  given redshift
bin:  $A$(redshifted  z=0.2-0.4  GOODS galaxies)  $\approx$  $A$(GOODS
galaxies) $<$  $A$(GOODS LIRGs) $<$ $A$(redshifted  local LIRGs). This
observed asymmetry  trend and Figure~\ref{COMP_UDF_GOODS}  (i.e., that
the observed asymmetry on  average correlates with intrinsic asymmetry
and  that  the  asymmetry  correction also  increases  with  intrinsic
asymmetry),  indicate  that the  asymmetry  corrections  based on  the
asymmetry deficits  of redshifted low-z  GOODS galaxies should  give a
reasonable estimate of the  true corrections for GOODS field galaxies.
The corrections for GOODS LIRGs could  be as small as the values based
on  redshifted low-z  GOODS galaxies  or as  large as  those  based on
redshifted local LIRGs. We therefore show these two cases as lower and
upper limits.  Note  that GOODS LIRGs with $L_{\rm  IR}$ $>$ 10$^{11}$
$L_{\odot}$ follow the same  relation, although they are slightly more
symmetric  than  LIRGs  with  $L_{\rm  IR}$  $>$  2.5$\times$10$^{11}$
$L_{\odot}$.  As the  asymmetry deficit  at a  given redshift  shows a
large     variation    for     different     low-z    objects     (see
Figure~\ref{asym_correction_LIRGs} and Figure~\ref{asym_corr_AllGAL}),
the whole  distribution of the asymmetry  deficit is used  to assign a
probability for the  final corrected asymmetry of a  galaxy at a given
redshift.  For field galaxies,  such an asymmetry deficit distribution
is further constructed as a function of the galaxy $B$-band brightness
(see Figure~\ref{asym_corr_AllGAL}).

\subsection{Evolution of the Merger Fraction in LIRGs}\label{Mor-Evo-LIRG}

We now correct  the observed asymmetry of GOODS  LIRGs using asymmetry
deficits  determined from  redshifted low-z  GOODS galaxies  and local
LIRGs,  which  give  conservative  lower- and  upper-limit  estimates,
respectively, as discussed above.   With these corrections, the merger
fraction of GOODS LIRGs as a function of redshift can be derived. Note
that the asymmetry criterion for mergers is usually defined as $A$ $>$
0.35  \citep[e.g.][]{Conselice03}.  Given a  systematic shift  of 0.05
between  our  asymmetry  method   and  that  in  the  literature  (see
\S~\ref{New_Noise}),    we    adopted    $A$    $>$    0.3    to    be
consistent\footnote{A small fraction of  galaxies with $A$ $>$ 0.3 are
  not true  major mergers, such  as the highly inclined  disk galaxies
  and      some      star-forming      galaxies     with      multiple
  non-symmetrically-distributed bright HII regions. Here, we adopt $A$
  $>$  0.3 as  a practical  definition of  galaxy mergers  and  do not
  correct  possible   contaminators.   }.   Figure~\ref{MergeFrac_Red}
shows our  result for LIRGs  at $L_{\rm IR}$  $>$ 2.5$\times$10$^{11}$
L$_{\odot}$ and  LIRGs at $L_{\rm IR}$ $>$  10$^{11}$ L$_{\odot}$. The
shaded areas  are enclosed  by lower- and  upper-limits to  the merger
fractions. The figure shows that  LIRGs are dominated by mergers up to
$z$ $=$ 1.2, with merging  fractions $\sim$ 50\% in all redshift bins.
Consistent   with  what  has   been  found   in  the   above  section,
high-redshift LIRGs  show slightly  lower merger fractions  than local
LIRGs.
  
As a test of the  conclusions from the asymmetry calculations, we also
visually   classified   the   IRAS   Revised  Bright   Galaxy   Sample
\citep{Sanders03}  as a function  of the  IR luminosity.   For $L_{\rm
  IR}$ $<$ 10$^{11.5}$ L$_{\odot}$, we used the Digital Sky Survey for
galaxies at  distance $<$ 60 Mpc  where the spatial  resolution is $<$
400  pc.  At 10$^{11.5}$  $L_{\odot}$ $<$  $L_{\rm IR}$  $<$ 10$^{12}$
$L_{\odot}$,  HST images were  used.  The  merger fractions  are 12\%,
41\%, 80\%  and 95\%, respectively, in IR  luminosity bins log($L_{\rm
  IR}$/$L_{\odot}$) of [10.5, 10.99],  [11.0, 11.49], [11.5, 12.0] and
[12.0,  12.49], where the  fraction for  the last  bin was  taken from
\citet{Sanders96}.   At $L_{\rm  IR}$ $<$  10$^{12}$  $L_{\odot}$, our
numbers  are consistent  with \citet{Sanders96},  lying  between their
pure merger fraction and  merger+close-pair fraction, as some galaxies
in close pairs have disturbed  morphologies and are thus identified as
mergers  by  us.   Multiplying  these  fractions  with  the  local  IR
luminosity functions, we obtained merger fractions of 45\% and 80\% at
$L_{\rm  IR}$   $>$  10$^{11}$   $L_{\odot}$  and  $L_{\rm   IR}$  $>$
10$^{11.5}$ $L_{\odot}$ as  shown in Figure~\ref{MergeFrac_Red}. These
values are nearly identical to the asymmetry-identified ones.

Our  result  of  high  merger  fractions  for  high-redshift  LIRGs  is
consistent  with the   UDF  result  \citep{Shi06}  where  no
corrections  are applied  for the  observed asymmetry  given  the much
deeper exposure in the UDF.  Such consistency further indicates that the
adopted correction  is valid  for the  high-redshift GOODS LIRGs.

We discuss briefly possible causes  for the low merger fractions found
in  other  studies:   \citet{Bridge07}  identified  mergers  based  on
asymmetry  parameters;  \citet{Zheng04}  and  \citet{Bell05}  visually
classified mergers; and \citet{Lotz08a} used Gini$-M_{20}$ to identify
mergers.  \citet{Bridge07} quantified the  merger fraction of LIRGs in
the  {\it Spitzer}  First  Look  Survey (FLS),  where  the imaging  is
shallower  than GOODS.   They did  asymmetry corrections  according to
\citet{Conselice05}, which we have found underestimate the correction
for  LIRGs  (see  \S~\ref{red-dep-asym-LIRG}).  This  under-correction
probably  accounts  for the  low  merger  values ($\sim$10-15\%)  they
derive.    \citet{Zheng04}  obtained   a   visually-classified  merger
fraction  of 16\%  for LIRGs  at  0.4$<$$z$$<$1.2 in  the CFRS  field.
Figure~\ref{LIRG_IMG_RED} warns of  the bias of visual classification,
as merging galaxies can look like normal galaxies with no or some weak
asymmetric    structures    due   to    SB    dimming.    Note    that
Figure~\ref{LIRG_IMG_RED} is constructed for  the GOODS field.  At the
depth  of the  images in  the CFRS  field, more  asymmetric structures
should  be lost  and  high-redshift LIRGs  will  be artificially  more
symmetric.    A  similar   bias   possibly  exists   for  the   visual
classification  used  in   \citet{Bell05}.   The  discrepancy  between
Gini$-M_{20}$  and  asymmetry  may  be  caused in  part  by the different
timescales over which galaxies can  be identified as mergers.  For gas
rich  galaxy mergers,  the timescale  for asymmetry  is  several times
longer than that for Gini-$M_{20}$ \citep{Lotz08b}.
 
However, we have compared the performance of CA and Gini-$M_{20}$ on
a local  sample of galaxies and  confirmed that to  first order they
give similar  results for  merger fractions. Since  Gini-$M_{20}$ is
less  affected  by the  ratio  of  S/N, it  is  not  clear what  the
implications of  our analysis would be for  morphology studies using
it  at $z\sim$1.  Although \citet{Lotz08a}  discuss these  issues, a
more detailed investigation would be desirable.
 
\subsection{Evolution of the Merger Fraction in Field Galaxies}\label{MOR-EVO-GAL}

\begin{figure*}
\epsscale{1.}
\plotone{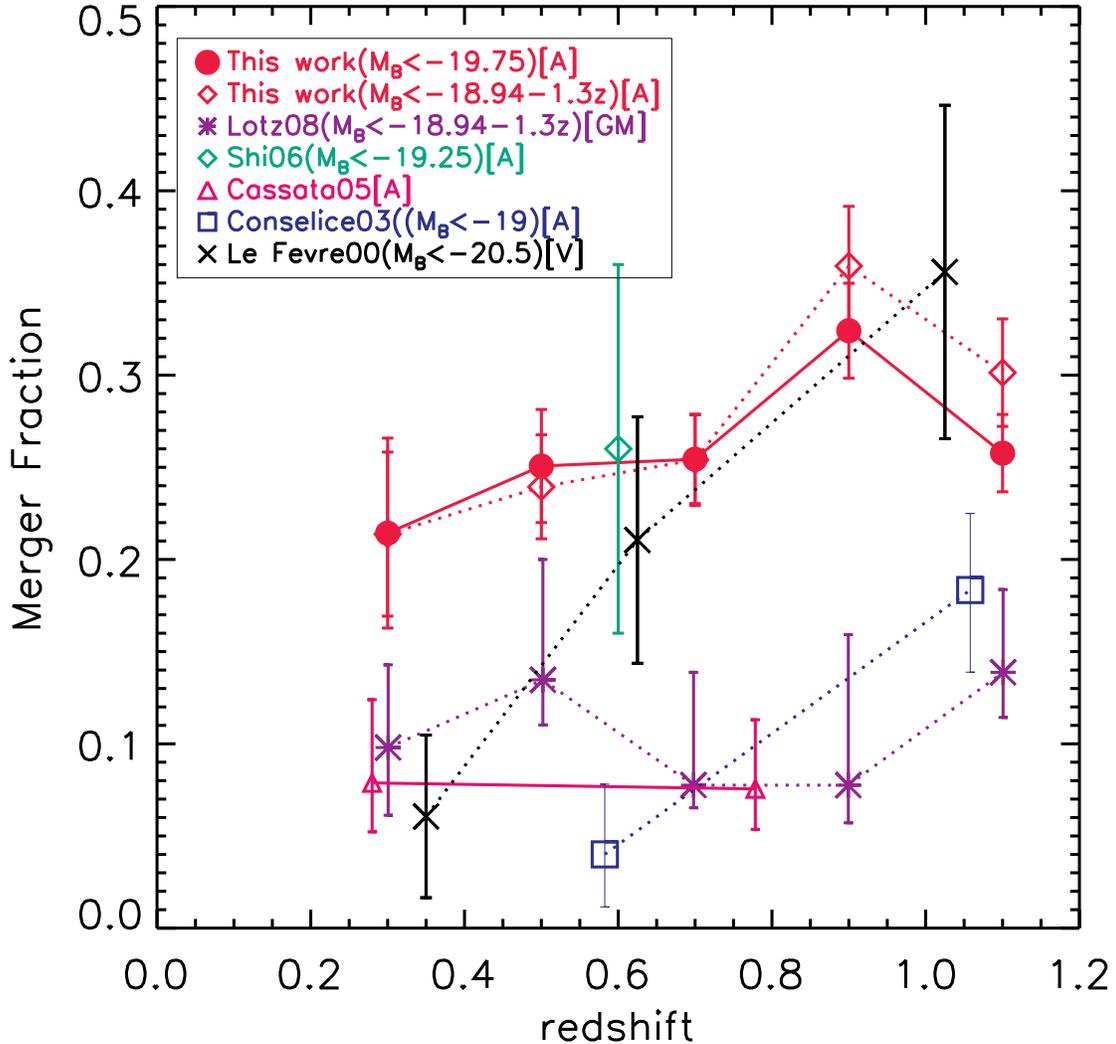}
\caption{\label{MergeFrac_Red_All} Redshift evolution of galaxy merger fractions with given $B$-band magnitude cuts. 
$''A''$: mergers  identified with  asymmetry; $''$GM$''$:
  Gini-$M_{20}$ classified mergers; $''V''$: visually classified mergers.}
\end{figure*}

Figure~\ref{MergeFrac_Red_All} shows the  merger fraction for galaxies
at $M_{B}$ $<$  -19.75 and $M_{B}$ $<$ -18.94-1.3z,  compared to other
works,  where  the asymmetry  correction  is  based  on the  asymmetry
deficits found for redshifted low-z  GOODS galaxies.  Our work shows a
weak redshift  evolution of galaxy merger  fractions, $f_{\rm merger}$
$\propto$  $(1+z)^{0.5\pm0.3}$  for  $M_{B}$  $<$ -19.75  and  $f_{\rm
  merger}$ $\propto$ $(1+z)^{0.9\pm0.3}$  for $M_{B}$ $<$ -18.94-1.3z.
This   result  is   consistent   with  the   morphological  study   of
\citet{Lotz08a}.   Again, however,  our merger  fraction  (20-30\%) is
several times higher than  other works except for \citet{Shi06}.  This
could  be  caused by  the  fact  that  the visual  classification  and
asymmetric classification  suffers from strong  redshift dependence as
discussed above.  Only \citet{Shi06} used images deep enough to detect
asymmetric features  as faint  as for nearby  galaxies and  thus their
result based on uncorrected asymmetry is consistent with our result.

We notice that our relatively-high merger fraction (20\%) at z=0.2-0.4
may require  a rapid evolution to  that at z=0.   However, the current
understanding of  the galaxy merger  fraction at $z<$0.2 is  much less
constrained mainly due  to the small volume and  use of shallow images
with  poor  spatial   resolution.   For  example,  \citet{DePropris07}
measured a merger fraction with asymmetry for a complete galaxy sample
from Millennium  Galaxy Catalog (MGC)  and found a merger  fraction of
2-4\% depending on the  definition of possible contaminators. However,
the  poor spatial resolution  (1.3 arcsec)  and shallow  exposure (sky
noise  = 26  mag  arcsec$^{-2}$) of  the  MGC survey  \citep{Cross04},
provide rest-frame  $B$-band image quality for galaxies  at z=0.1 only
comparable to  z=1 galaxies observed  in the GOODS survey,  implying a
possibly significant  under-estimate of the merger fraction  due to SB
effects.   The image quality  is even  worse for  the SDSS  and 2dFGRS
images. Current  {\it HST} data  do not provide a  statistically large
complete  galaxy   sample  at  $z$$<$0.2.  The   fraction  of  mergers
identified  as galaxy  pairs is  also not  well constrained,  from 1\%
\citep{DePropris07} to 5\% \citep{Lin08} for  the same set of data but
with  different  methods,  which  again suggests  that  more  thorough
studies are  required to  have good constraints  on galaxy  mergers at
$z$$<$0.2.

\subsection{Infrared and $B$-band Luminosity Functions of Galaxy Mergers}\label{IRLF-MERGERS}

\begin{figure*}
\epsscale{1.2}
\plotone{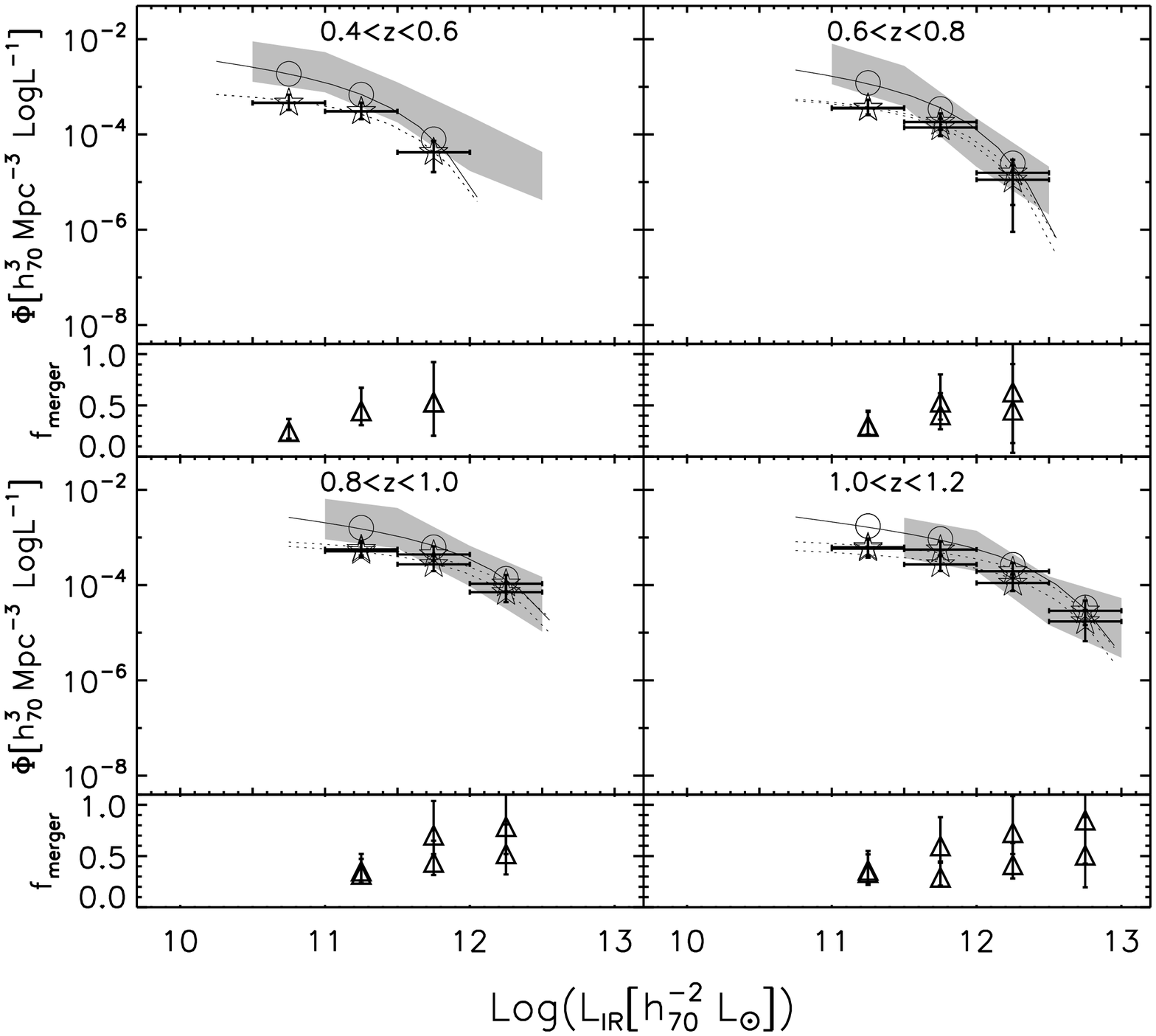}
\caption{\label{IRLF_mergers} Infrared  luminosity functions of galaxy
mergers  at  different redshifts  (open  stars), where  upper- and lower-limits
  at a given  IR luminosity correspond  to the result with 
asymmetry corrections based on redshifted local LIRGs and redshifted low-z GOODS galaxies
for galaxies with $L_{\rm IR}$ $>$ 2.5$\times$10$^{11}$ $L_{\odot}$,
respectively, while  asymmetry corrections based  on redshifed low-z
GOODS galaxies are  used for galaxies with lower  IR luminosity. The
open  circles are the  infrared luminosity  function of  GOODS field
galaxies. The grey area  shows the luminosity function and 3$\sigma$
uncertainty  of field  galaxies obtained  by  \citet{LeFloch05}. The
dotted line  is the fit  to the IR  LF of galaxy mergers,  while the
solid line is the fit to the IR LFs of general field galaxies.  The merger fraction
as a function of the IR luminosity is also shown. }
\end{figure*}

\subsubsection{Methodology for Constructing IR Luminosity Function}

The IR luminosity functions of galaxy mergers are derived
broadly following the works of \citet{LeFloch05} and \citet{Shi08}.
The 1/$V_{\rm max}$ method \citep{Schmidt68} is applied to galaxy mergers
with $m_{z}$ $<$ 25, $f_{24{\mu}m}$ $>$ 60 $\mu$Jy and known redshifts.
The comoving number density in a given luminosity bin can be written as:
\begin{equation}
\Phi({\rm Log}L_{\rm IR})d({\rm Log}L_{\rm IR}) = {\sum}\omega/V_{\rm max}
\end{equation}
where $\omega$ is the weight to correct for incompleteness and $V_{\rm max}$ is
the maximum volume for an object to be included in the sample. $V_{\rm max}$
is given by
\begin{equation}   
V_{\rm   max}    =   \int_{z_{\rm   low}}^{z_{\rm high}} \Omega \frac{dV}{dz} dz,
\end{equation} 
where [$z_{\rm low}$, $z_{\rm high}$]  is the redshift over which the
object can  be detected and $\Omega$  is the survey  solid angle (730
arcmin$^{2}$). Note  that the field edge excluded  for the morphology
study is  a small  fraction (1\%) of  the total area.   While $z_{\rm
  low}$ is  always fixed to the  low end of a  redshift interval, the
maximum redshift, $z_{\rm high}$ is defined as:
\begin{equation} 
z_{\rm high}={\rm min}(z_{\rm bin}^{\rm high}, z^{\rm limit}_{\rm IR},
                         z^{\rm limit}_{m_{z}}),
\end{equation} 
where $z_{\rm bin}^{\rm high}$ is the high end of a redshift interval,
and $z^{\rm  limit}_{\rm IR}$  is the limiting  redshift at  which the
observed  IR  flux reaches  the  limiting  flux  of 60  $\mu$Jy.   The
K-correction for 24 $\mu$m flux is based on the star-forming templates
of \citet{Rieke09}.  $z^{\rm  limit}_{m_{z}}$ is the limiting redshift
where the  observed $z$-band  magnitude reaches $m^{\rm  limit}_{z}$ =
25. The  K-correction in the $z$-band  is based on  the ACS photometry
and KCORRECT code \citep{Blanton03, Blanton07}.

The incompleteness correction $\omega$ includes corrections for the IR
detection, IR objects associated  with optical counterparts at $m_{z}$
$<$ 25,  the redshift  measurement success rate  and the  criterion of
secure  morphologies.  The incompleteness  of the  IR detections  as a
function of the 24 $\mu$m flux density is given in \citet{Papovich04}.
Our  sample is limited  to $f_{24{\mu}m}$  = 60  $\mu$Jy at  which the
incompleteness is  $\sim$50\%.  There is  no incompleteness correction
for IR  objects associated with optical counterparts  with $m_{z}$ $<$
25.  This  is because $m_{z}$ $<$  25 is deep enough  to detect nearly
all   optical   counterparts  of   the   IR   objects  brighter   than
$f_{24{\mu}m}$ =  60 $\mu$Jy, given the rough  correlation between the
IR  luminosity  and  optical  luminosity  in  \citet{LeFloch05}.   The
incompleteness correction for redshift  measurements is defined as the
ratio of  the number of all  objects to that of  objects with redshift
measurements  within a  three dimensional  magnitude-color-color space
$m_{z}$-$(B-i)$-$(V-z)$.   Such corrections  can account  for redshift
measurement success as a function  of the galaxy brightness and color.
For the  whole ACS  photometry sample at  $m_{z}$ $<$ 25,  the success
rate for  redshift measurements is  68\%. The final correction  is for
galaxies with  secure morphologies, i.e., galaxy  radius $R_{\rm gal}$
$>$  15 pixel  and  asymmetry  uncertainty error(A)  $<$  0.1. Such  a
correction is  defined as the  ratio of the  number of all  objects to
that of objects with secure morphologies within a given luminosity bin
and  a given  redshift bin.   Note  that these  corrections are  small
($<$1.2) except for the highest luminosity  bin in 0.4 $<$ $z$ $<$ 0.6
(a  correction factor  of  1.5).   This indicates  that  the final  IR
luminosity function of galaxy mergers  is mainly based on objects with
secure morphologies.

\subsubsection{IR Luminosity Function of Galaxy Mergers}

Figure~\ref{IRLF_mergers} shows the  IR luminosity functions of galaxy
mergers within different  redshift bins.  As shown in  the figure, our
IRLFs of  field galaxies match  those of \citet{LeFloch05}  well.  The
1-$\sigma$ uncertainties  of the IRLFs  of galaxy mergers  include the
Poisson  noise,  the   uncertainty  in  the  24$\mu$m-to-$L_{\rm  IR}$
conversion and  cosmic variance.   Since our field  size is  almost as
large as that  in \citet{LeFloch05} and 24$\mu$m flux  density is used
to derive the  total IR luminosity in both  works, for the uncertainty
in  the 24$\mu$m-to-$L_{\rm  IR}$ conversion  and cosmic  variance, we
simply adopted the result obtained by \citet{LeFloch05}, an average of
0.2 dex upper-side uncertainty and 0.1 dex lower-side uncertainty. The
final  uncertainty   including  the   Poisson  noise  is   plotted  in
Figure~\ref{IRLF_mergers} and listed in Table~\ref{IRLF_MERGERS}.  The
solid lines  are the  fit to the  IR LFs  of field galaxies  while the
dotted lines  are the fit to  the IRLFs of galaxy  mergers.  The curve
for the fit is the Schetcher function. Although the Schetcher function
fails to  fit the local  field galaxy IR  LF \citep{Perez-Gonzalez05},
our limited  available data points  do not allow  us to use  a formula
with more  parameters, such  as a double  power-law. The slope  of the
Schetcher function is fixed as the  average value of the fit to the IR
LFs in  the first two  redshift bins. The  fitted result is  listed in
Table~\ref{IRLF_MERGERS_FIT}.

Figure~\ref{IRLF_mergers} also shows the fraction of galaxy mergers as
a function  of the  IR luminosity in  different redshift bins.  At all
redshifts,  the merger fraction  increases with  the IR  luminosity, a
trend similar to that for local galaxies \citep{Sanders96}.

\begin{figure*}
\epsscale{1.}
\plotone{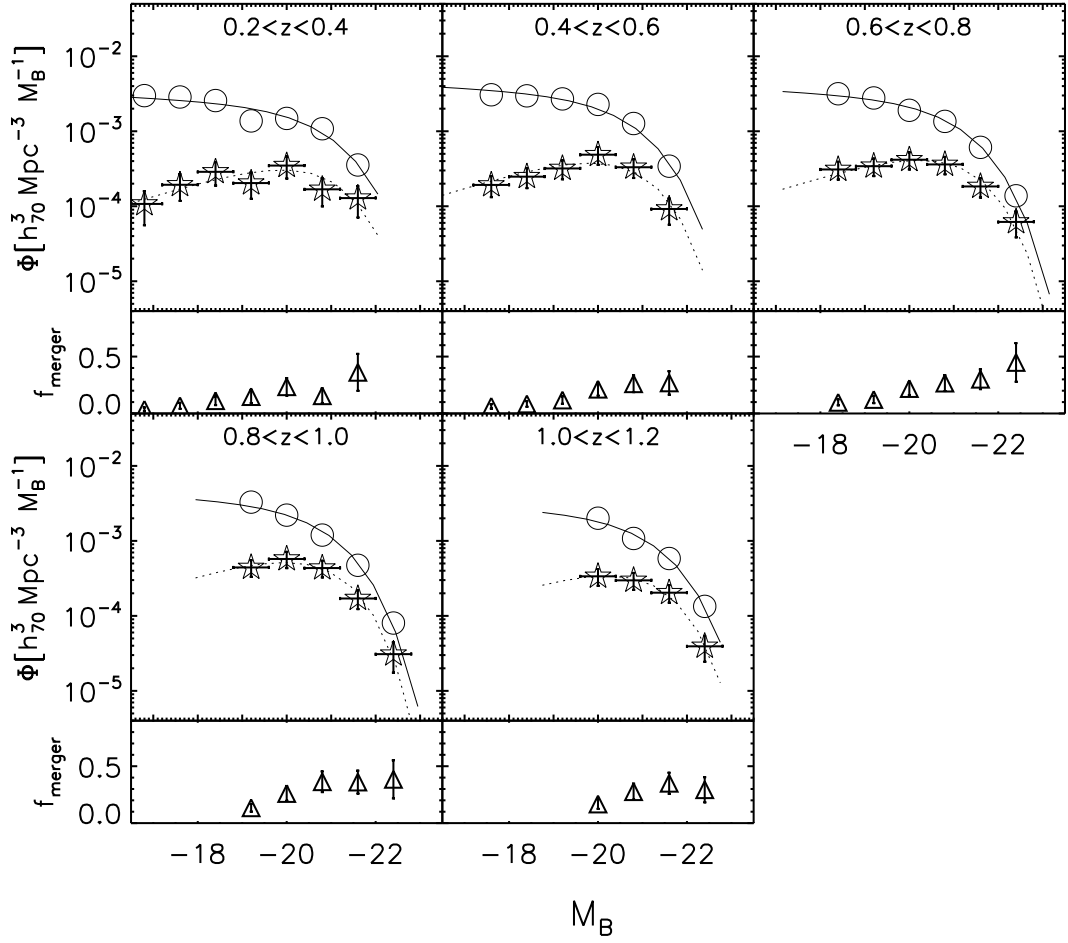}
\caption{\label{MbLF_mergers} $B$-band luminosity functions of galaxy mergers (open stars) and general 
field galaxies (open circles) at different redshifts. The dotted line is the fit to the LF of 
galaxy mergers while the solid line is the fit to that of general field galaxies. }
\end{figure*}

\begin{figure}
\epsscale{1.}
\plotone{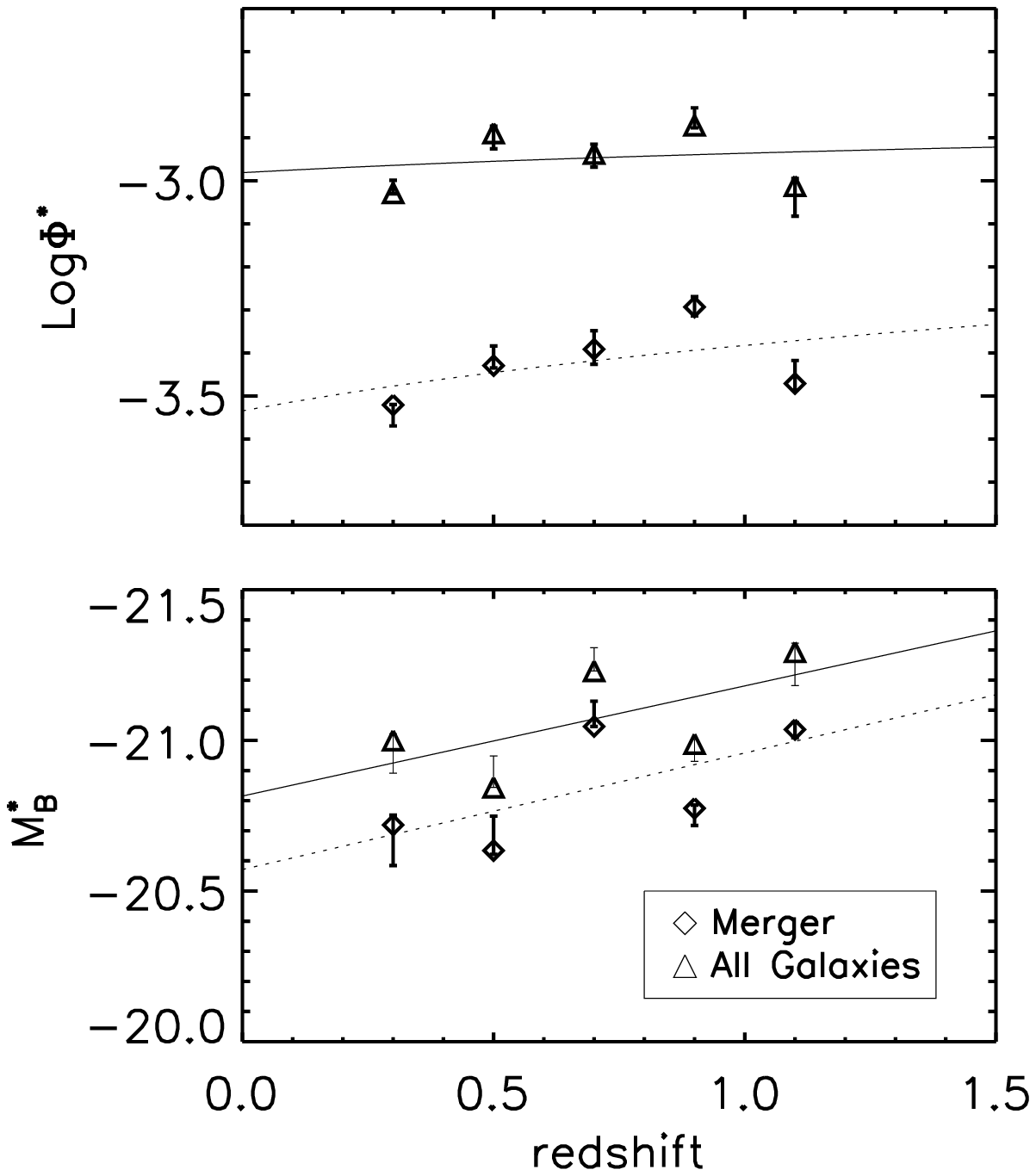}
\caption{\label{EVO_DEN_LUM_MbLF} Evolution in the luminosity and density of $B$-band luminosity functions of galaxy mergers (diamonds) and general 
field galaxies (triangles). }
\end{figure}

\subsubsection{$B$-band Luminosity Function of Galaxy Mergers}

The  rest-frame $B$-band  luminosity functions  of galaxy  mergers are
also derived as shown  in Figure~\ref{MbLF_mergers}.  The result  of the  merger $B$-band  LF is  listed in
Table~\ref{MbLF_MERGERS}.  Again, Schetcher  functions are used to fit
the data points  where the slope is fixed as the  average value of the
fit to the LF in the  first three redshift bins.  The fitted result is
listed  in  Table~\ref{MbLF_MERGERS_FIT}.  For the  B-band  luminosity
functions,   there    are   enough   data   points    to   break   the
luminosity-density    degeneracy.    The    result    is   shown    in
Figure~\ref{EVO_DEN_LUM_MbLF}.  For  the   general  field  B-band  LF,
$\Phi^{*}{\propto}(1+z)^{0.1\pm0.5}$                                and
$\Delta$$M_{B}^{*}{\propto}(-0.4\pm0.3){\Delta}z$, which are generally
consistent   with  the   literature  data   \citep[see][and  references
  therein]{Faber07}.       For     the      merger      B-band     LF,
$\Phi^{*}{\propto}(1+z)^{0.5\pm0.5}$                                and
$\Delta$$M_{B}^{*}{\propto}(-0.4\pm0.3){\Delta}z$.

The breaking of the luminosity-density degeneracy allows us to
evaluate the evolution of the pure merger fraction:
\begin{equation}
n_{\rm galaxy}f_{\rm merger} = n_{\rm merger}
\end{equation}
Given $n_{\rm galaxy}$${\propto}(1+z)^{0.1\pm0.5}$ and $n_{\rm merger}$${\propto}(1+z)^{0.5\pm0.5}$ in the $B$-band, we have
\begin{equation}
f_{\rm merger}^{B} \propto (1+z)^{0.4\pm0.7}
\end{equation}

The  weak  redshift evolution  of  the  pure  merger fraction  $f_{\rm
  merger}^{\rm B}$ indicates that the  observed weak evolution in a given
$B$-band magnitude  is really caused  by the weak  evolution of  the galaxy merger
number  density relative to the total number density. The merger rate can be estimated through
the merger fraction and   the  timescale  of 0.2-1.1  Gyr  over which  the
asymmetry  parameter  can  identify  mergers \citep{Lotz08b}.

%%%%%%%%%%%%%%%%%%%%%%%%%%%%%%%%%%%%%%%%%%%%%%%%%%%%%%%%%%%%%%%%%%%%%%%%%%%%%%%%%%%%%%%%%%%%%%%
\section{Discussion}\label{Discussion}

\subsection{Merger-Dominated High-Redshift LIRG Morphologies and High Merger Fractions In General Field Galaxies}\label{Mer-Dom-LIRG}

While the local LIRGs  are dominated by mergers \citep{Sanders96}, the
result for high-redshift LIRGs is controversial, with merger fractions
from  as low  as  10-20\% \citep{Zheng04,  Bell05, Bridge07,  Lotz08a,
  Melbourne08} to  $\sim$50\% \citep{Shi06}.  Motivated  by the result
of \citet{Shi06}  in the UDF field,  we re-evaluated the  effect of SB
dimming  on  the  morphology   measurements.   We  first  revised  the
asymmetry  measurement based  on  simulations.  Our  new method  shows
a smaller scatter and does not  suffer from a systematic offset compared
to  the   one  in  the   literature.   We  then   obtained  redshift-,
IR-luminosity- and  optical-luminosity-dependent asymmetry corrections
by  measuring  asymmetry  deficits  of  local LIRGs  and  low-z  GOODS
galaxies redshifted to different  higher redshifts.  By applying these
corrections, we  have found that high-redshift LIRGs  are generally
highly asymmetric  galaxies with implied merger  fractions $\sim$ 50\%
up to $z$=1.2.  For the  general galaxy population, we obtained a weak
evolution   of   merger   fractions,   consistent  with   results   of
\citet{Bundy04}, \citet{Lin08}, \citet{Lotz08a} and \citet{deRavel08},
but with several times higher merger fractions (20-30\%).

In  \S~\ref{Mor-Evo-LIRG}   and  \S~\ref{MOR-EVO-GAL},  we  summarized
possible reasons for the lower fractions of morphologically-identified
mergers  in  other  studies  \citep{LeFevre00,  Zheng04,  Conselice03,
  Cassata05, Bell05, Bridge07, Lotz08a, Melbourne08}.

The low  fraction of  kinematic galaxy pairs  \citep{LeFevre00, Lin08,
  deRavel08} is at least partly due to the different timescales probed
by this  method compared to morphology studies.  Also, kinematic pairs
are  exclusively  major  mergers  while  a small  fraction  of  highly
asymmetric objects may be minor  mergers.  Our high merger fraction is
consistent with the incidence  of mergers identified through disturbed
velocity  fields.  Specifically,  studies  of the  velocity fields  of
$\sim$60 galaxies  at 0.4 $<$  $z$ $<$ 0.75  \citep{Neichel08, Yang08}
find a  low fraction of rotationally-supported disk  galaxies and high
fraction   (41$\pm$7\%)   of   galaxies   with   complex   kinematics.
\citet{Neichel08}  also show  that  the auto  classifications (CA  and
Gini-M$_{20}$  without corrections)  miss  half of  the galaxies  with
complex  kinematics  and mis-classify  them  as  normal disk  galaxies
\citep[also     see     a     similar    result     from     numerical
  simulations;][]{Lotz08b}.

\subsection{Contribution by Galaxy Mergers to the Cosmic IR Energy Density}\label{Evo-Mer-IR}

\begin{figure}
\epsscale{1.}
\plotone{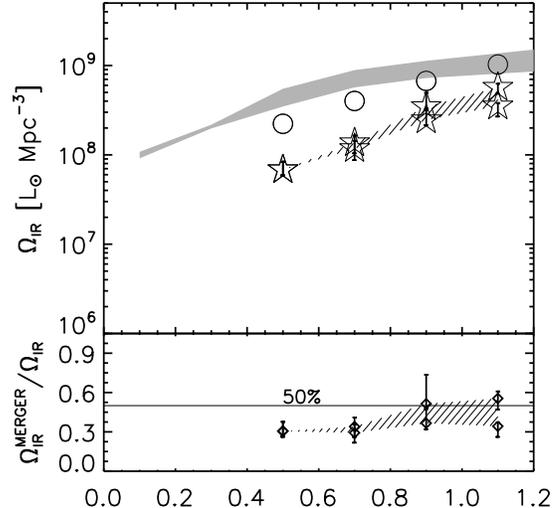}
\caption{\label{LFden_mergers} Top Panel:  The total cosmic IR density
(open   circles)   compared   to   the   result   (grey   area)   of
\citet{Perez-Gonzalez05}.  The open star  symbols are the upper- and
lower-limits of  the cosmic  IR density with  1-$\sigma$ uncertainty
contributed  by   galaxy  mergers  (see  Figure~\ref{IRLF_mergers}).
Bottom panel: The ratio and  $1-\sigma$ uncertainty of the merger IR
density to the total IR density. }
\end{figure}

We have derived IR LFs of galaxy mergers at different redshifts out to
$z$=1.2.  With  these IR LFs,  we now can quantitatively  evaluate the
contribution of galaxy mergers to  the cosmic IR energy density out to
$z$   $\sim$   1.2.   The   open   circle   in   the  top   panel   of
Figure~\ref{LFden_mergers}   shows  the   cosmic  IR   energy  density
estimated  by  integrating  general  IR  LFs obtained  in  this  work,
compared  to the  work of  \citet{Perez-Gonzalez05} (grey  area).  The
open star is  the evolution of the cosmic IR  energy density of galaxy
mergers obtained by integrating the IR LFs of galaxy mergers where the
upper  and  lower limits  correspond  to  the  result with  asymmetry
corrections  based on  redshifted local  LIRGs and  low-z  GOODS field
galaxies,  respectively.  The  bottom panel  of the  figure  shows the
ratio of the merger IR energy density to the total IR energy density.

The galaxy  mergers-produced IR energy density shows dramatic evolution:
$\Omega_{\rm  IR}^{\rm   mergers}$  $\propto$  (1+$z$)$^{\sim5}$  and
(1+$z$)$^{\sim6}$  for  the  result   with  low  and  high  asymmetry
corrections, respectively,  compared to the total  IR energy density
evolution  (1+$z$)$^{3.0}$ \citep{LeFloch05,  Perez-Gonzalez05}.  At
$z$  $>$ $\sim$1 the cosmic  IR  energy density  is dominated  by
galaxy mergers.   This result can  be expected since  LIRGs dominate
the   IR   energy  density   at   $z$   $>$  0.7   \citep{LeFloch05,
  Perez-Gonzalez05} and  LIRGs are  dominated by galaxy  mergers, as
found in this work.

To  convert the  cosmic energy IR density  to  the cosmic  SFR density,  the
unobscured  star formation as  traced by  the UV energy  density needs  to be
included.  In the local universe, the  UV energy density corresponds to about 30\% of the
cosmic   SFR  density   and   decreases   to  10\%   at   $z$  =   1.2
\citep{LeFloch05}.  The indicated correction from the  cosmic IR energy density  to the total
SFR density for non-mergers should be larger than those for mergers, as
the  merger IR energy density has contributions from galaxies  with on-average
higher IR luminosity. This implies  that Figure~\ref{LFden_mergers}  may over-estimate the
contribution of galaxy mergers to the total cosmic SFR density.

\section{Conclusions}\label{conclusions}

We present detailed morphology studies of intermediate-redshift (0.2$<z<$1.2) LIRGs and general
field galaxies in the GOODS field by measuring galaxy concentration
and asymmetry, with the goal to constrain the role of galaxy mergers
in the cosmic star formation history. Our main conclusions are as follow:

(1) A  new asymmetry determination  method is proposed to  account for
the  importance  of   underlying  background  structures  in  accurate
asymmetry measurements.  Simulations indicate that our method does not
suffer from a systematic offset and has small intrinsic uncertainty.

(2)  The   redshift  dependence  of   the  galaxy  asymmetry   due  to
surface-brightness dimming is a function of the asymmetry itself, with
larger corrections for more asymmetric objects.  This requires careful
asymmetry corrections for high-redshift galaxies.

(3) With the necessary  asymmetry corrections, high-redshift LIRGs are
generally galaxies with high  asymmetries and have implied merger fractions
of $\sim$50\% up to z=1.2, although they are slightly more symmetric than local LIRGs.

(4)  With similar asymmetry  corrections, high-redshift  general field
galaxies  show a  weak redshift  evolution of  merger fractions  up to
$z$=1.2 but with a relatively high merger fraction (20-30\%).

(5) The  B-band luminosity functions of galaxy  mergers show evolution
in the  luminosity and density  similar to general field  galaxies. By
removing the luminosity-density degeneracy, the pure number density of
galaxy  mergers relative  to  the total  density  shows weak  redshift
evolution, i.e., $(1+z)^{m}$ with $m=0.4\pm0.7$.

(6)  The IR  luminosity functions  of  galaxy mergers  are derived  in
several redshift  bins: they indicate that the  merger fraction increases
with  the  infrared  luminosity.   The integral  of  these  luminosity
functions shows that the cosmic  IR energy density from galaxy mergers
shows a  dramatic redshift  evolution ($(1+z)^{\sim5\textendash6}$) and  starts to
dominate the total cosmic IR energy density at $z>$$\sim$1.

Our  study  has   been  confined  to  the  CA   method  of  morphology
determination. The  significant corrections we derive  for this method
indicate that  a similar study of other  methods (e.g., Gini-$M_{20}$)
would be desirable.

\acknowledgements

We  thank the anonymous  referee for  detailed comments.   Support for
this work was provided by NASA through contract 1255094 issued by JPL/
California Institute of Technology.  JML acknowledges support from Leo
Goldberg  Fellowship  sponsored  by  the  National  Optical  Astronomy
Observatory.   P.G.   P.-G.   acknowledges  support from  the  Spanish
Programa Nacional  de Astronom\'{\i}a y  Astrof\'{\i}sica under grants
AYA 2006--02358  and AYA 2006--15698--C02--02, and from  the Ram\'on y
Cajal  Program financed  by the  Spanish Government  and  the European
Union.

%\appendix

\section{ Appendix: New Noise Correction}\label{New_Noise}

We revisit here  the problem of the sky noise  asymmetry correction.  In  the case of high redshift  galaxies, we find that
using  the   minimum  noise  asymmetry  to   correct  the  min($A_{\rm
  galaxy+noise}$),  as proposed by  \citet{Conselice00}, overestimates
the true  galaxy asymmetry  and results in  a relatively  large error.

\begin{figure}
\epsscale{0.7}  
\plotone{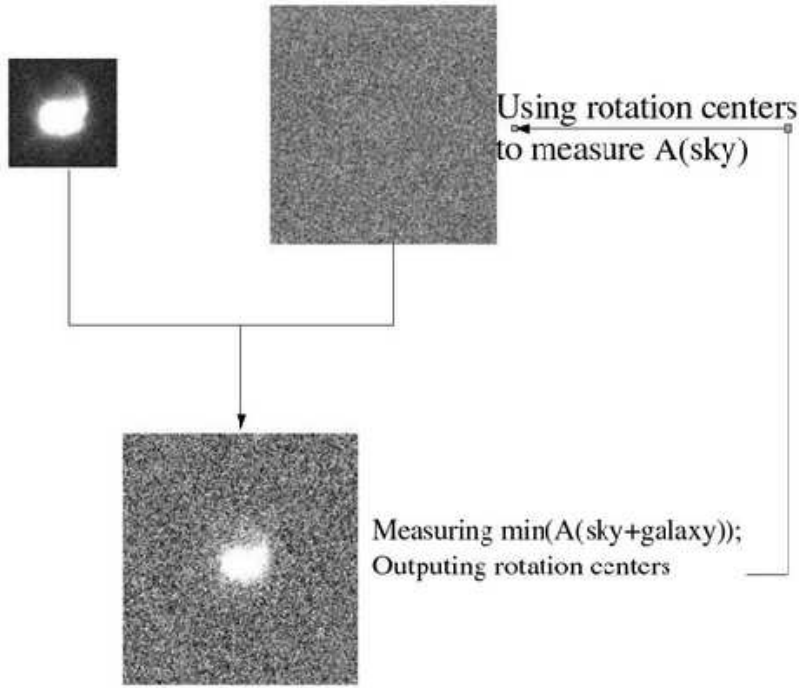}
\caption{\label{Simu_Illustrate} 
This  cartoon illustrates  the basic  idea for  finding the  real noise
correction for ${\rm min}(A_{\rm  galaxy+noise})$.  A UDF galaxy image
is superposed on a clean GOODS background region to create an artificial GOODS
galaxy.   For this  galaxy,  the galaxy  size,  concentration and  the
rotation  centers  giving  ${\rm  min}(A_{\rm galaxy+noise})$  can  be
measured.   The resulting rotation  centers and  galaxy size  are then
used to  measure the  asymmetry of the  clean GOODS background  region, which
should  give   the  real   noise  correction  for   ${\rm  min}(A_{\rm 
galaxy+noise})$.}
\end{figure}

The noise  correction is critical  for an accurate measurement  of the
true  galaxy  asymmetry.  For  example,  even  for  the minimum  noise
asymmetry correction  (i.e., $A_{\rm noise}^{\rm  corr}$ = min($A_{\rm
  noise}$),  referred to  as  the minimum  method  in the  following),
$A_{\rm  noise}^{\rm  corr}$  is  a significant  fraction  of  $A_{\rm
  galaxy+noise}$.   The median  value of  min($A_{\rm noise}$)/$A_{\rm
  galaxy+noise}$ reaches almost 70\%  for the local relatively high SB
galaxy sample from \citet{Frei96}, which \citet{Conselice00} used as a
test sample for the asymmetry  parameter. We have used the GOODS-South
field and the UDF to test ways to measure asymmetry.  We carried out a
simulation to optimize the estimate of the
true   $A_{\rm  noise}^{\rm   corr}$   underlying  ${\rm   min}(A_{\rm
  galaxy+noise})$, using noise asymmetries measured in fields placed randomly around the galaxy.  The
basic     idea    of    the     simulation    is     illustrated    in
Figure~\ref{Simu_Illustrate}.  An UDF galaxy image was superposed on a
clean GOODS  background region to  create an artificial  GOODS galaxy.
For  this galaxy,  the  galaxy size,  concentration  and the  rotation
centers giving  ${\rm min}(A_{\rm galaxy+noise})$  were measured.  The
resulting  rotation center  and galaxy  size were  then used  with the
galaxy removed to measure the  asymmetry of the clean GOODS background
region,  which should give  the true  $A_{\rm noise}^{\rm  corr}$.  By
putting circular regions with the  same aperture size randomly on this
background region, the  randomly-produced noise asymmetry distribution
was  obtained.   The  goal  was  to  find the  best  way  to  use  the
randomly-produced  noise asymmetry distribution  to estimate  the true
$A_{\rm   noise}^{\rm  corr}$.    In  this   experiment,   the  direct
measurements  of the  true noise  correction  values let  us test  our
method.

Here we  describe the  details of the  simulation.  First we  used the
GOODS South  $z$-band catalog to select  a total of  13 clean $z$-band
GOODS   background  regions  with   sizes  of   400$\times$400  pixels
(1 pixel=0.03$''$).  As the GOODS  catalog only indicates if the galaxy
center is outside the region or  not, we inspected each region to make
sure there was no extended emission from galaxies with centers outside
the region.   These regions were  distributed over the whole  field to
account  for any variations  over the  field.  We  then picked  25 UDF
galaxies spanning  a range  of S/N (50-1000)  within half  radii.  The
main reason for  using UDF galaxies is that the UDF  is deep enough to
detect faint asymmetric structures.  Our  test will thus not be biased
by  low S/N.   The UDF  segmentation images  were used  to  define the
galaxy pixels.  These galaxy pixels  were superposed on each of the 13
GOODS background  regions.  For each  UDF galaxy and  GOODS background
region, 121 artificial  images were created by putting  the UDF galaxy
at each position of a 11$\times$11 grid with a separation of 10 pixels
over the central 100$\times$100  pixel regions of the GOODS background
image.  For these  121 artificial images, we measured  the galaxy size
and  true  noise  asymmetry  correction  as  described  in  the  above
paragraph. We then used the mean size of these galaxies to produce the
distribution of random noise asymmetry corrections by putting circular
regions randomly  within the GOODS  background image.  Based  on these
randomly-produced corrections within  the average aperture, we created
1000 random corrections for  each artificial galaxy by multiplying the
corrections within the average  aperture with the aperture size (area)
of  the  artificial  galaxy  relative  to the  average  aperture  size
\citep{Conselice00}.   In summary, for  one UDF  galaxy and  one GOODS
background  image,  we  have  121  true  noise  asymmetry  corrections
corresponding  to  121 artificial  galaxies  and  for each  artificial
galaxy  we  have  a   distribution  of  1000  randomly-produced  noise
asymmetry  measures.  Note  that  the noise  associated  with the  UDF
galaxy itself is not an issue in this experiment, as any noise pattern
associated with the UDF galaxy acts like the galaxy signal in terms of
searching    for   rotation    centers   giving    ${\rm   min}(A_{\rm
  galaxy+noise})$.

\begin{figure}
\epsscale{0.6}
\plotone{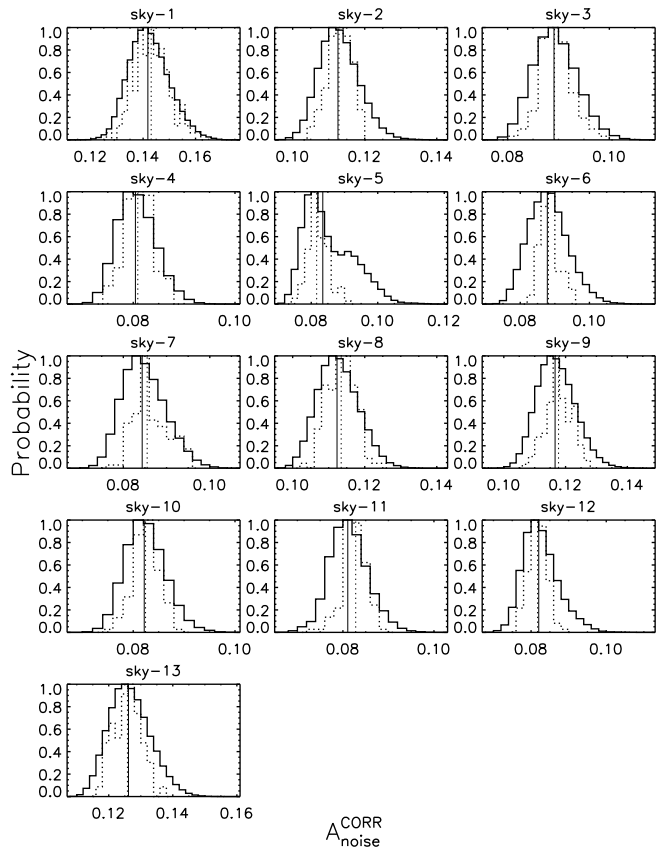}
\caption{\label{comp_real_rad_dis_SN_1000} Comparisons  between  the
true noise asymmetry corrections (dotted line; 121 asymmetry values in
each  panel) and  the randomly-produced  noise asymmetry  (solid line;
121$\times$1000 asymmetry values  in each panel) for 13  clean GOODS background
regions  superposed with  a high  S/N ($\sim$1000)  UDF  galaxy. The vertical lines
are the median values of two distributions.}
\end{figure}

\begin{figure}
\epsscale{0.6}
\plotone{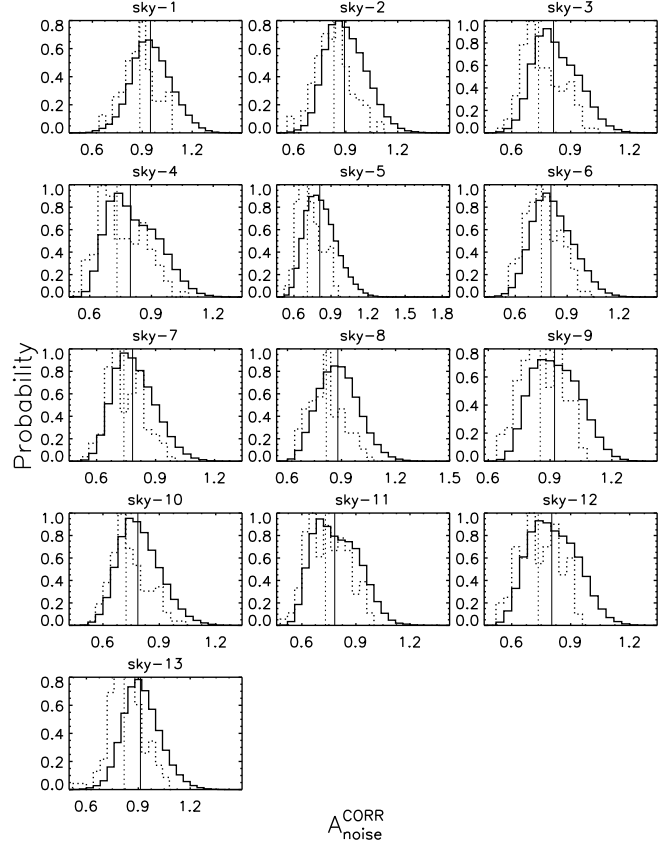}
\caption{\label{comp_real_rad_dis_SN_70} The same as in Figure~\ref{comp_real_rad_dis_SN_1000} but
for a low S/N (70) UDF galaxy.} 
\end{figure}

Figure~\ref{comp_real_rad_dis_SN_1000}  shows  comparisons  between  the
true noise asymmetry corrections (dotted line; 121 asymmetry values in
each  panel) and  the randomly-produced  noise asymmetry  (solid line;
121$\times$1000 asymmetry values  in each panel) for the 13  clean GOODS sky
regions  superposed with  a high  S/N ($\sim$1000)  UDF  galaxy, while
Figure~\ref{comp_real_rad_dis_SN_70}   shows  result   for  a   low  S/N
($\sim$70) UDF  galaxy.  The vertical line indicates  the median value
of   each   distribution.   Figure~\ref{comp_real_rad_dis_SN_1000}   and
Figure~\ref{comp_real_rad_dis_SN_70}  indicate that  using  the minimum
noise  asymmetry  to  correct  the noise  asymmetry  underlying  ${\rm
  min}(A_{\rm  galaxy+noise})$  will   overestimate  the  real  galaxy
asymmetry in  many cases.  The true sky  asymmetry correction actually
lies approximately  between two extreme  cases, the minimum  value and
the median of the randomly-produced noise asymmetry.  The minimum case
is  reached when the  galaxy is  so faint  that the  ${\rm min}(A_{\rm
  galaxy+noise})$  is actually dominated  by the background.  The
median correction is appropriate when the galaxy is so bright that the
${\rm  min}(A_{\rm  galaxy+noise})$ is   dominated by  galaxy
signal,   as  shown   in   Figure~\ref{comp_real_rad_dis_SN_1000}.  
Because we usually apply  a lower limit S/N cut, we will
rarely have the former case.   A more important implication from these
two figures is that the  backgrounds always have unknown structures, which
makes accurate noise correction impossible.  Even for the same UDF
galaxy and  the same GOODS background  image, the scatter  of the true
noise asymmetry  correction is significant. For a  high S/N  (1000) UDF
galaxy as  shown in Figure~\ref{comp_real_rad_dis_SN_1000},  the scatter
is   about  20\%   (or  $\sigma$$(A)=$0.02)   and it reaches   50\%  (or
$\sigma$$(A)=$0.2)  for  a  low  S/N  (70)  UDF  galaxy  as  shown  in
Figure~\ref{comp_real_rad_dis_SN_70}.    The  difference  in   the  true
correction  between different GOODS  background regions  superposed with
the same  UDF galaxy is  also non-negligible. The fluctuation  in the
background  can be caused  by non-uniform  exposure, variation  in the
detector response,  Poisson noise and probably more  important for the
deep survey, a large number of objects below the detection limit.

\begin{figure}
\epsscale{0.8}
\plotone{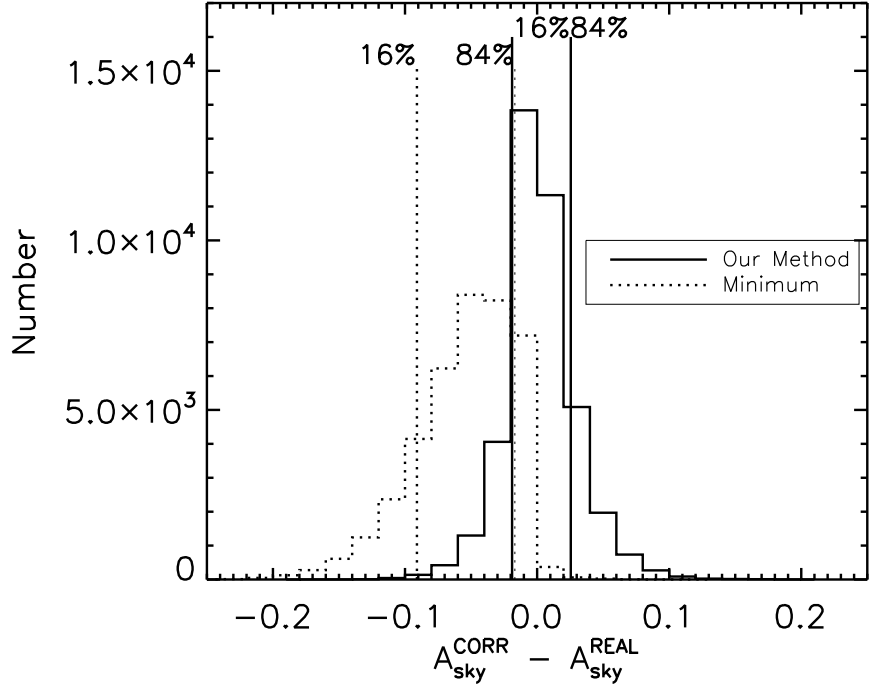}
\caption{\label{comp_min_ourmethod} The solid histogram shows the
difference between the noise correction of our 15\% probability method and the true
correction, while the dotted one shows the difference between the correction of the minimum method
and the true correction. Vertical lines indicate the 16\% and 84\% 
low-end probability tail of these two histograms. The median offsets of our method and
the minimum one are 0.00 and -0.05, respectively. The 68\% confidence range of our method
and minimum one are 0.044 and  0.074, respectively.}
\end{figure}

Given the significant scatter  in the true  noise correction  caused by
complicated  background  structures and  the  fact  that  we can  only
measure the  noise asymmetry in the background  region around (instead
of underlying)  the galaxy,  we  tried to  find the  best way to  use the
randomly-produced noise corrections to estimate the true value.  To do
this, we  measured the  fraction of the  randomly-produced corrections
below    the     true    value    for     each    artificial    galaxy.
The   median   value  of the distribution of this fraction
is    0.15. Therefore, the value at the 15\% probability low-end tail
of the  randomly-produced noise corrections gives the  best estimate of
the true value.   Figure~\ref{comp_min_ourmethod} compares our new noise
correction (described as the 15\% probability method in the  following) to
the minimum method where the  minimum correction is simply the minimum
of    the     randomly-produced    corrections.     As     shown    in
Figure~\ref{comp_min_ourmethod},  the   minimum-noise  corrected  galaxy
asymmetry  overestimates  the true  value,  with  a  median offset  of
${\delta}A$=-0.05. It  may not be a  severe problem to  use the minimum
noise correction as  long as all the morphologies  are measured in the
same way.  However, we emphasize that, as discussed above and shown in
Figure~\ref{comp_real_rad_dis_SN_1000},
Figure~\ref{comp_real_rad_dis_SN_70}  and Figure~\ref{comp_min_ourmethod},
it is impossible  to recover an accurate noise  correction due to the
complicated   background   structures.   Figure~\ref{comp_min_ourmethod}
indicates that the 68\% confidence range of $A_{\rm noise}^{\rm corr}$
- $A_{\rm noise}^{\rm real}$ of our  method is 0.044 compared to 0.074
for the minimum method. A scatter of 0.074 is actually quite large given
the  merging  criteria  of  $A$   $>$  0.35  for  the  minimum  method
\citep{Conselice03}.   Our 15\% probability method  provides  almost  two  times  less
scatter in the $A_{\rm noise}^{\rm corr}$ - $A_{\rm noise}^{\rm real}$
compared to the minimum method.

\begin{figure}
\epsscale{0.7}
\plotone{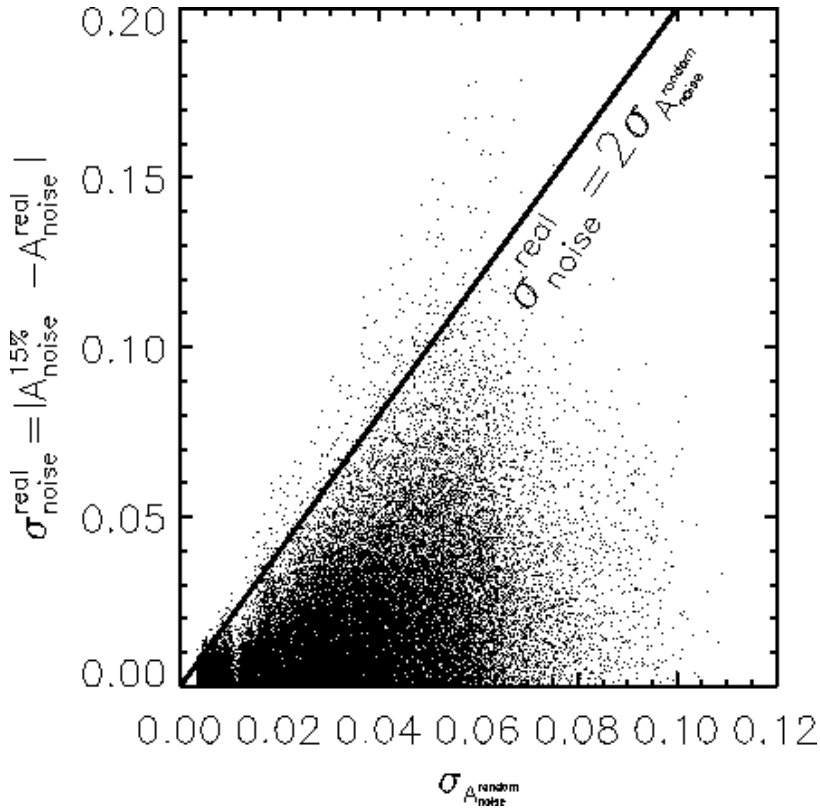}
\caption{\label{error_noise_corr}  The   difference   ($\sigma_{\rm
  noise}^{\rm real}$ =  $|A_{\rm noise}^{\rm 15\%}$-$A_{\rm noise}^{\rm
  real}|$) between the true  correction and our 15\% probability method
as the  function of  the standard deviation  ($\sigma_{\rm noise}^{\rm
  random}$) of  the randomly-produced noise  corrections. The solid line
shows $\sigma_{\rm
  noise}^{\rm real}$=2$\sigma_{\rm noise}^{\rm
  random}$ which can be used to provide conservative estimate of $\sigma_{\rm
  noise}^{\rm real}$ for the 99\% objects in the simulation.}
\end{figure}

Figure~\ref{error_noise_corr}   shows  the   difference  ($\sigma_{\rm
  noise}^{\rm real}$ = $|A_{\rm noise}^{\rm 15\%}$-$A_{\rm noise}^{\rm
  real}|$) between the true correction and our 15\% probability method
as  a function  of  the standard  deviation ($\sigma_{\rm  noise}^{\rm
  random}$)  of  the randomly-produced  noise  corrections.  As  shown
above,  the  noise  corrections  are affected  by  unknown  background
structures and thus it would  be expected that there is no correlation
between $\sigma_{\rm  noise}^{\rm real}$ and  $\sigma_{\rm noise}^{\rm
  random}$.  However, the  maximum $\sigma_{\rm noise}^{\rm real}$ for
a given  $\sigma_{\rm noise}^{\rm random}$ is  roughly correlated with
$\sigma_{\rm  noise}^{\rm  random}$.   As  shown by  the  solid  line,
2$\times$$\sigma_{\rm noise}^{\rm random}$  should give an upper limit
to $\sigma_{\rm  noise}^{\rm real}$ valid  for 99\% of the  objects in
the simulation (e.g. effectively at about 3$\sigma$ significance).

As a summary,  we propose a new noise asymmetry  correction.  A set of
1000  randomly-produced  noise  corrections  is  produced  by  putting
circular regions in the background  image around the target. The value
of this distribution  at the 15\% probability low-end  tail is used to
correct the noise asymmetry for min($A_{\rm galaxy+noise}$). The error
($\sim$3$\sigma$) in  the final measured galaxy asymmetry  is taken to
be two  times the standard deviation of  these randomly-produced noise
asymmetries. In reality, some galaxies are always present in the field
of targets. To  account for this problem, we  defined the success rate
for  the one-thousand circular  region placements  as the  fraction of
circular  regions  containing  no   galaxy  signal  indicated  by  the
SExtractor segmentation image.  Circular regions containing any galaxy
signal were not used.  More  sets of one-thousand placements were done
until  one  thousand successful  measurements  were  reached.  If  the
success  rate  for one  set  of placements  is  lower  than 50\%,  the
circular region size for the following set is taken to be 80\% of this
set.  Then the measured background  asymmetry is rescaled to that with
the original size by assuming background asymmetry proportional to the
aperture area \citep{Conselice00}.

\section{Morphologies of Redshifted Local LIRGs and Low-z GOODS Field Galaxies}\label{Red-Dep-Gal-Asy}
\subsection{Morphologies of Redshifted Local LIRGs}

\begin{figure}
\epsscale{0.7}
\plotone{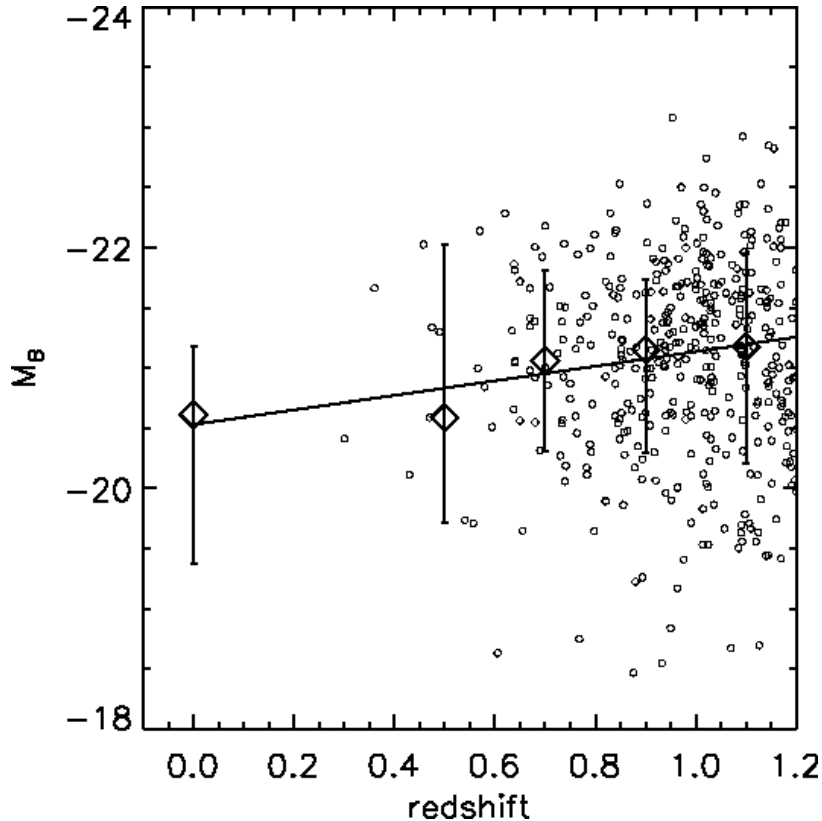}
\caption{\label{MB_LIRG}  The  redshift   evolution  of  the  $B$-band
magnitude of  LIRGs with $L_{IR}$ $>$  10$^{11.4}$ L$_{\odot}$, where
the z=0 point is for  local LIRGs and higher-redshift points are for
LIRGs in GOODS.  Diamonds with error bars show the median values and
68\%  confidence ranges  of the  $B$-band magnitude  distribution at
$z$=0,   0.4$<$$z$$<$0.6,   0.6$<$$z$$<$0.8,   0.8$<$$z$$<$1.0   and
1.0$<$$z$$<$1.2.  The solid  line  gives the  linear  fit $m_{B}$  =
-20.53-0.60$z$. }
\end{figure}

The  $B$-band morphologies  of the  local  LIRGs at  $L_{\rm IR}$  $>$
10$^{11.4}$   L$_{\odot}$  were   measured  following   the  procedure
described in \S~\ref{Revised_Asy}.  Due to the high resolution and the
nature of complicated morphologies, about 10\% of the local LIRGs have
multiple  radii at  $\eta(r)$=0.2.   For these  objects,  we used  the
maximum radius  at $\eta(r)$=0.2 that  matches the visually-identified
galaxy size well.

To account  for the redshift-dependence of LIRG  morphologies, we also
measured  the  redshifted  $B$-band  morphologies of  these  LIRGs  at
redshifts of  0.5, 0.7,  0.9 and 1.1.   Since we quantified  the GOODS
galaxy morphologies  using the rest-frame $B$-band images,  we put the
local LIRGs  into a  clean GOODS $V$-band  sky region for  local LIRGs
redshifted to z=0.5, $i$-band sky region for redshifted LIRGs at z=0.7
and z=0.9, and $z$-band sky  region for redshifted LIRGs at z=1.1. The
exposure times  for GOODS $V-$, $i-$  and $z-$ band  images are around
5000, 7000 and  20000 secs, respectively.  The clean  GOODS sky region
has a  size of  24$''$$\times$24$''$, which is  large enough  to cover
very extended tidal  tails.  The GOODS field is too  crowded to have a
continuous  24$''$$\times$24$''$ clean  region  without any  intruding
object.  Our  field is  obtained by merging  four 12$''$$\times$12$''$
clean regions  whose sky fluctuations  are consistent with  each other
within 3\%.

To  redshift  the LIRGs,  the  pixel  size  was rebinned  through  IDL
FREBIN.pro which rebins the pixel  assuming the flux within a pixel is
constant over  the pixel area.  We  did not apply  any size evolution,
since on average the galaxy size  $R_{\rm p}$ of LIRGs does not change
with  redshift.  The  DN/sec of  the  local LIRG  $B$-band images  was
converted  to the  GOODS ACS  counts at  a given  band by  using their
PHOTOFLAM values  and then decreased  by the square of  the luminosity
distance  and  (1+$z$),  where  (1+$z$)  is  the  k-correction,  since
PHOTOFLAM is defined as inverse  sensitivity in units of erg cm$^{-2}$
s$^{-1}$  $\AA^{-1}$.   The resulting  DN/sec  is  then brightened  by
0.60$z$ magnitudes, which accounts  for the redshift dependence of the
rest-frame    $B$-band    magnitude    of    LIRGs   as    shown    in
Figure~\ref{MB_LIRG}.

During  this pixel-rebinned  and flux-rescaled  process,  the original
LIRG image noise was correspondingly scaled down.  To mimic the galaxy
morphologies  measured in  the  real GOODS  sky  region, the  original
galaxy  background  fluctuation should  not  dominate  over the  GOODS
background in measuring the  redshifted local LIRGs.  To quantify when
the scaled  original galaxy background does not  affect the redshifted
LIRG  morphology measurement,  we carried  out a  test to  measure the
asymmetry of a clean GOODS sky region superposed on a second GOODS sky
region   scaled  down   by  a   given  factor.  When  the second  sky  region has  a
scaled-down  factor   of  $<$  0.3,   its  effect  on   the  asymmetry
measurements  of the  first sky  region is  negligible.  Therefore, we
measured all redshifted LIRGs  with rescaled original background noise
smaller than 0.3 of the GOODS sky noise. At redshifts of 0.5, 0.7, 0.9
and 1.1, the number of measurable redshifted LIRGs are 61, 83, 88 and 88
out of a total of 88  local LIRGs. Therefore, the redshifted LIRGs are
still representative of the complete local LIRG sample.

\subsection{Morphologies of Redshifted Low-z Field Galaxies}\label{red-dep-asym-NORMGAL}

The galaxy  sample that we  used is the  $V$-band images of  596 GOODS
galaxies at 0.2  $<$ $z$ $<$ 0.4 with  secure morphology measurements.
For  a   fiducial  galaxy  at  z=0.3,  the   GOODS  $V$-band  exposure
corresponds to a 10$\sigma$  rest-frame $B$-band surface brightness of
25.5  mag   arcsec$^{-2}$,  comparable  to   the  depth  of   the  HST
observations of  local LIRGs (25.0 mag  arcsec$^{-2}$).  The algorithm
to measure morphologies  of redshifted galaxies is almost  the same as
for redshifted  local ULIRGs  with two differences.  First we  did not
require the  rescaled original  galaxy noise to  be $<$ 0.3  times the
noise in the  band where galaxies are redshifted  to.  Instead we used
the following strategy:
\begin{equation}
I_{\rm new} = (1-\eta^{2})^{0.5}B_{\rm new} + I_{\rm old}
\end{equation}
where  $I_{\rm new}$  is the  final new  image, $I_{\rm  old}$  is the
rebinned  and  rescaled original  galaxy,  $B_{\rm  new}$  is the  new
background image  in the  band where the  galaxy is redshifted  to and
$\eta$ is the ratio of  rescaled and rebinned original galaxy noise to
the fluctuation in $B_{\rm new}$.  This method will make sure that all
redshifted galaxies  will be used  at all redshifts  while maintaining
the noise level.  The main reason  that we did not use this method for
local  LIRGs  is  that  we  wanted  to  quantify  the  effect  of  the
complicated  background structures  in  deep fields  on the  asymmetry
measurements.

\begin{figure}
\epsscale{0.65}
\plotone{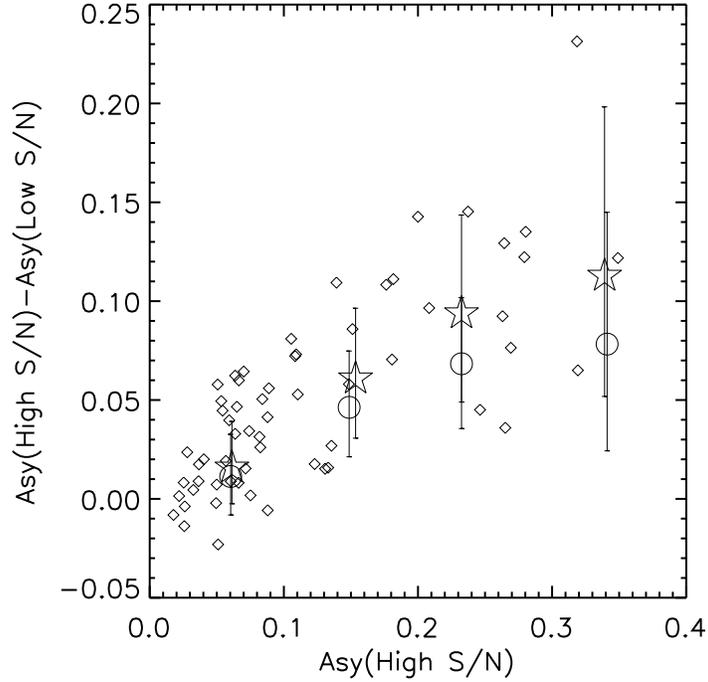}
\caption{  \label{COMP_UDF_GOODS_REDSHIFTGAL}  The  asymmetry  deficit
vs.  intrinsic  asymmetry.  The  open  circles  and  stars  are  for
low-redshift   (0.2$<$$z$$<$0.4)   GOODS   galaxies  redshifted   to
0.4$<$$z$$<$0.8 with $B$-band luminosity evolution and no evolution,
respectively.   Diamonds show  the
evolution-independent  result,  which is  the  asymmetry
difference using  GOODS images (low  S/N) and UDF images  (high S/N)
for galaxies with 0.4$<$$z$$<$0.8.  The  error bars are given at the
16\% and 84\% probability limits.}
\end{figure}

\begin{figure}
\epsscale{0.6}
\plotone{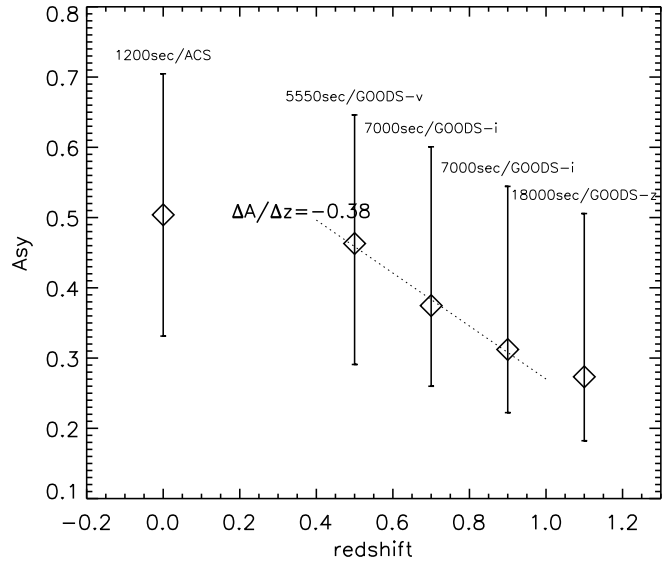}
\caption{\label{LocalLIRG_AsyRedshift}  The median asymmetry and its
68\% confidence  range for  local LIRGs and  redshifted local  LIRGs at
z=0.5,  0.7,  0.9 and  1.1.  The exposure  time  at  each redshift  is
labelled. A  linear fit  to
the central   three    points   with   similar    exposure   times   gives
${\delta}A/{\delta}z$=-0.38.}
\end{figure}

Second,  no  additional  $B$-band  luminosity  evolution  is  applied,
although there  is a luminosity  evolution in the  $B$-band luminosity
function  \citep[e.g.][]{Faber07}.   This  is  because,  ideally,  the
asymmetry  correction for  a distant  galaxy  should be  based on  the
asymmetry deficit of a  redshifted low-z galaxy with intrinsically the
same asymmetry and  $B$-band brightness.  Applying $B$-band luminosity
evolution will  artificially underestimate  the effect of  SB dimming.
To  further  demonstrate  this  argument, we  measured  the  asymmetry
deficits  for redshifted galaxies  with $B$-band  luminosity evolution
\citep[$\Delta$$M_{B}$   =   -1.1$\Delta$$z$;   ][]{Faber07}  and   no
evolution. This result can be  compared to the real result obtained by
comparing the  asymmetries using GOODS  images and UDF images  for the
same      galaxies.        The      result      is       shown      in
Figure~\ref{COMP_UDF_GOODS_REDSHIFTGAL}      for     galaxies     with
0.4$<$$z$$<$0.8 within which range the UDF limiting SB in the $z$-band
is  comparable to  the  GOODS  $v$-band limiting  SB  at $z$=0.3.   As
indicated by  the figure, the  cases with evolution  underestimate the
asymmetry correction significantly.

\subsection{The Asymmetry Deficit due to SB Dimming is a Function 
of the Asymmetry Itself}\label{red-dep-asym-LIRG}

\begin{figure}
\epsscale{0.8}
\plotone{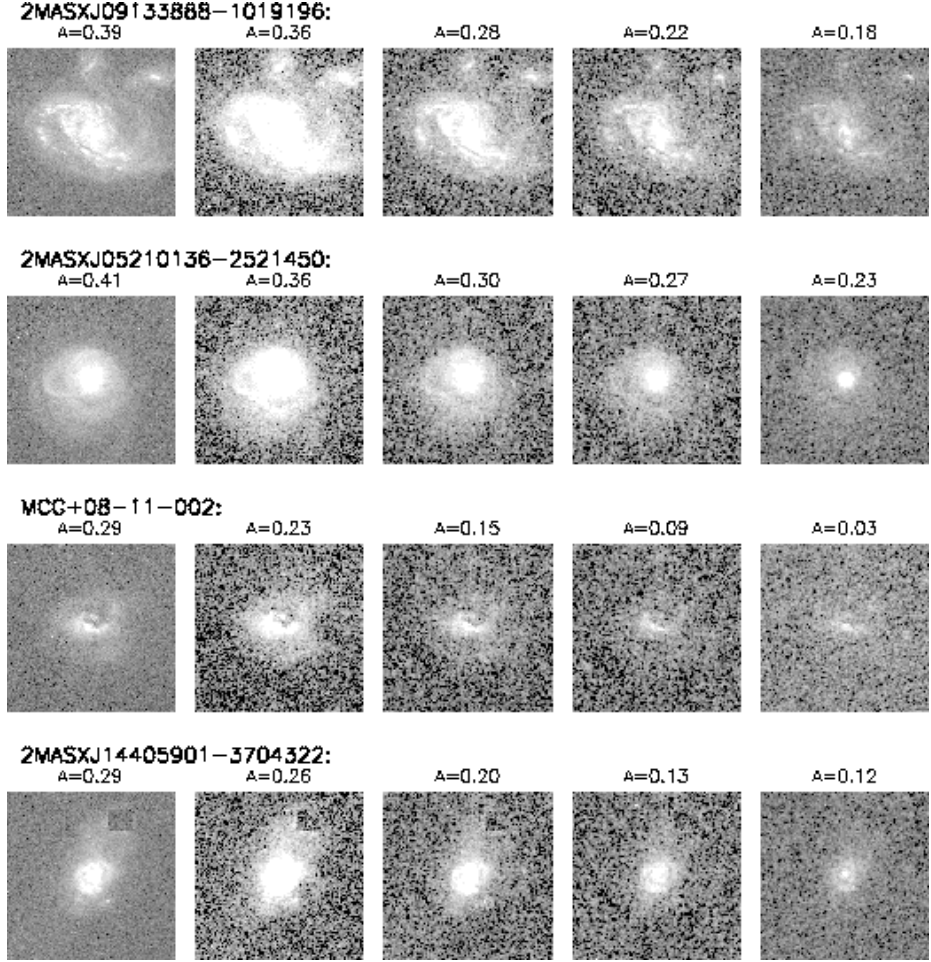}
\caption{\label{LIRG_IMG_RED}  Illustrations that apparent galaxy  morphologies
become    less   asymmetric    at   higher    redshift    as   the
low-surface-brightness  asymmetric structures  are lost  due  to
surface-brightness dimming.  From left to right, we  show the local
LIRGs  and  redshifted local  LIRGs  at  z=0.5,  0.7, 0.9  and  1.1,
respectively. The  image size is about  30$\times$30 kpc$^{2}$. }
\end{figure}

Figure~\ref{LocalLIRG_AsyRedshift}  shows the  median  asymmetries and
the 68\%  confidence ranges for  the local LIRGs and  redshifted local
LIRGs at z=0.5, 0.7, 0.9 and  1.1.  The exposure time at each redshift
is labelled. As shown in the figure, the galaxy appears more symmetric
at high redshifts as more low-SB asymmetric structures are embedded in
the  background fluctuations  due  to  SB dimming.   A  linear fit  to
central  three  points with  similar  exposure times  ($\sim$6000secs)
gives ${\delta}A/{\delta}z$=-0.38.   The asymmetry of  the local LIRGs
is  below the  extrapolation of  this linear  fit as  the  local LIRGs
should be  more asymmetric at the  exposure time of  $\sim$ 6000 secs.
The redshifted LIRGs at z=1.1  are above the extrapolation of this fit
as  a  deeper  exposure   at  z-band  detects  more  faint  asymmetric
structures.  Although the redshift  dependence of the galaxy asymmetry
has been noticed by \citet{Conselice00, Conselice03, Conselice05}, the
slope of our dependence is  much larger than theirs.  Even compared to
the test  sample of local irregular  galaxies \citep{Conselice05}, our
slope is $\sim$2 times larger.  We believed that the reason for such a
large  difference  is  the  different  local galaxy  samples  used  to
quantify the  redshift-dependence of the galaxy  asymmetry.  While the
test galaxy  sample in  Conselice's works is  a reasonable  option for
study  of  the  general   high-redshift  galaxy  population,  it  will
underestimate  the galaxy  asymmetry for  studies such  as quantifying
merger  fractions.   Figure~\ref{LIRG_IMG_RED}  illustrates how  tidal
tails  are progressively lost  and galaxies  appear more  symmetric at
higher redshift.

To further  illustrate the dependence of the  galaxy asymmetry deficit
due  to  SB  dimming  on  the asymmetry  itself,  we  measured  galaxy
asymmetries using  GOODS $z$-band images  and UDF $z$-band  images for
about   $\sim$250    UDF   galaxies   with   high    S/N   and   large
size. Figure~\ref{COMP_UDF_GOODS} clearly shows  that there is a trend
of a larger  asymmetry deficits due to SB  dimming for more asymmetric
galaxies.

\begin{figure}
\epsscale{0.7}
\plotone{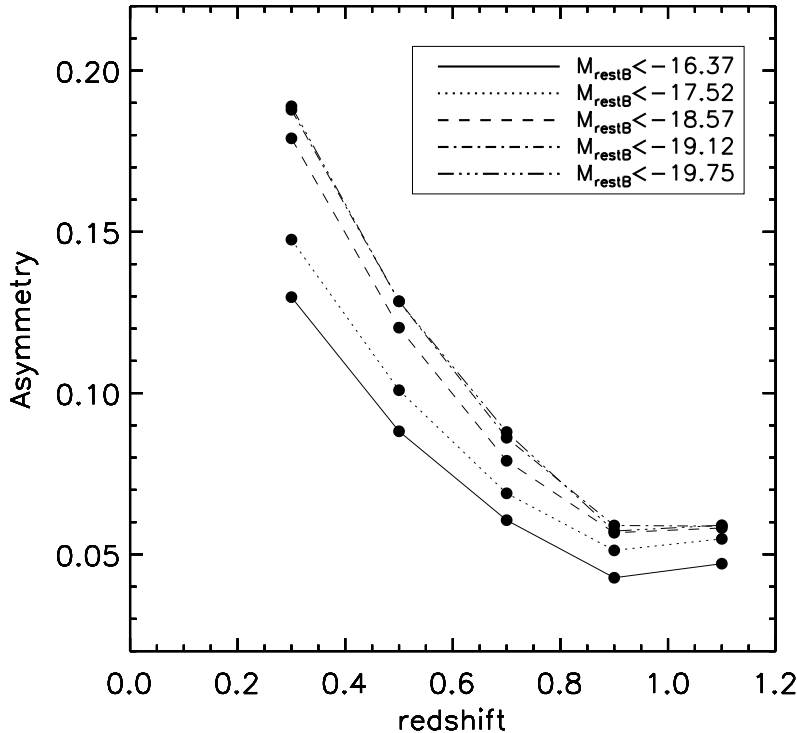}
\caption{\label{goods_z0204_redshifted} The median asymmetry of GOODS-S galaxies 
at 0.2 $<$ $z$ $<$ 0.4 and brighter than certain $B$-band magnitudes redshifted
to higher redshifts. }
\end{figure}

Figure~\ref{goods_z0204_redshifted}  shows the  asymmetry  deficit for
redshifted  GOODS galaxies at  0.2$<$$z$$<$0.4.  The  magnitude limits
correspond to the 70\% completeness of the redshift-morphology catalog
at     different      redshifts     (see     Figure~\ref{MB_LIR_red}).
Figure~\ref{goods_z0204_redshifted} further strengthens the conclusion
based on local LIRGs  that the redshift-dependence of galaxy asymmetry
is a  function of the  asymmetry itself. Brighter GOODS  galaxies with
higher  asymmetry at  $z$=0.3 show  a  steeper decline,  while even  the
brightest sample shows a slower decline than the local LIRGs that have
higher asymmetry.

\subsection{Asymmetry Deficits of Redshifted Local LIRGs and Low-z GOODS Field Galaxies}\label{asym-correction-detail}

\begin{figure}
\epsscale{0.65}
\plotone{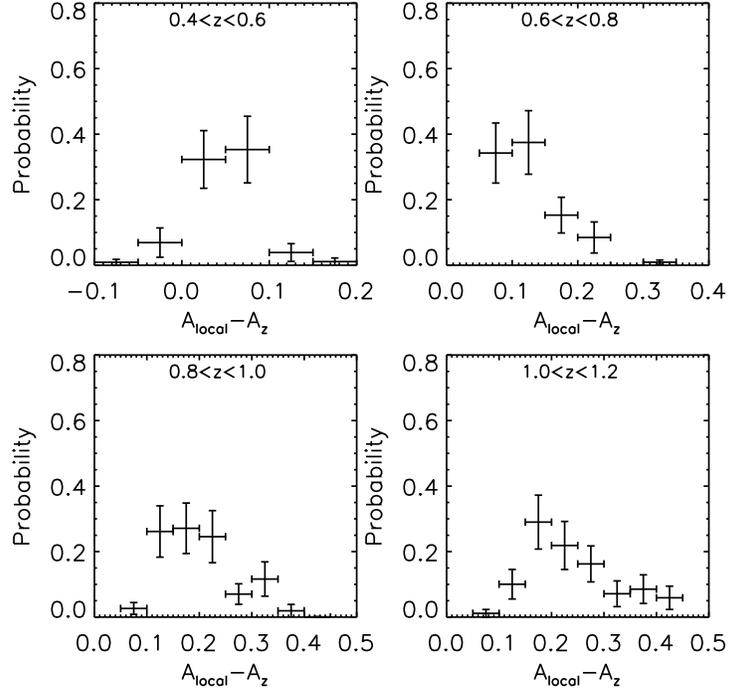}
\caption{\label{asym_correction_LIRGs}  The distribution of the asymmetry deficit for local LIRGs at 
$L_{\rm IR}$ $>$ 2.5$\times$10$^{11}$ L$_{\odot}$
redshifted to higher redshifts.}
\end{figure}

\begin{figure}
\epsscale{0.65}
\plotone{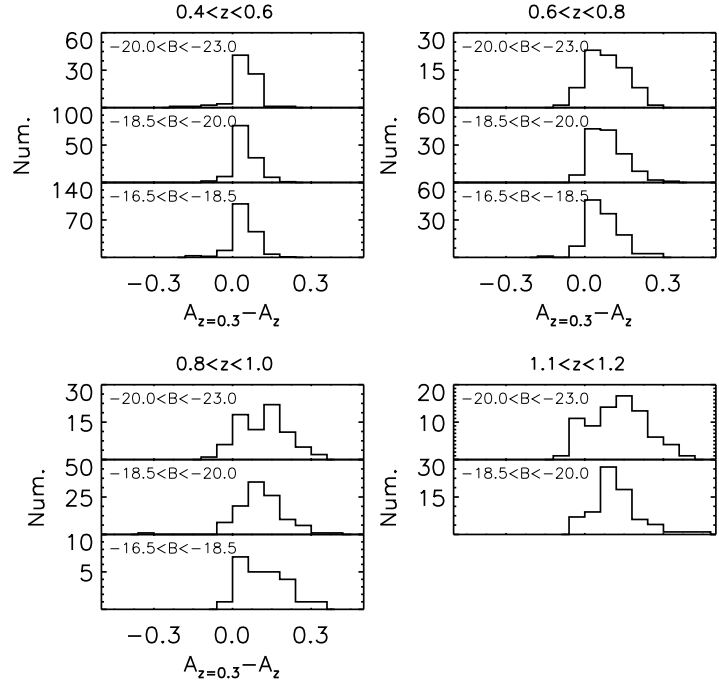}
\caption{\label{asym_corr_AllGAL}  The asymmetry deficit distribution of low-redshift (0.2$<$$z$$<$0.4) GOODS galaxies
redshifted to higher redshifts. The histograms are given in different $B$-band luminosity bins as labelled in the figure.}
\end{figure}

Figure~\ref{asym_correction_LIRGs} shows the probability distributions
of  the  asymmetry deficits  for  redshifted  local  LIRGs, where  the
uncertainty  is the  Poisson noise.   Note  that the  weight has  been
applied  for this flux  limited sample  to obtain  a volume-equivalent
result (see  \S~\ref{Gal-Sam-LIRG}).  The  bin size for  the asymmetry
distribution is 0.05, the  mean quadratic asymmetry error.  The median
value  of the  distribution in  a given  redshift bin  can be  used to
correct  LIRGs at  that redshift.   However, as  shown in  the figure,
there is a significant width in the distribution, implying a variation
in the asymmetry deficit for an individual galaxy. Such variations are
expected given the complicated  structures of galaxy morphologies as a
function of the SB.

For general galaxy populations, Figure~\ref{asym_corr_AllGAL} shows the
probability  distributions   (in  absolute  object   numbers)  of  the
asymmetry  deficit for  low-redshift (0.2$<$$z$$<$0.4)  GOODS galaxies
redshifted  to  higher redshifts  as  functions  of absolute  $B$-band
magnitude. The  bin size for  the asymmetry distribution is  0.06.  At
the  highest  redshift, the  distribution  for  the faintest  $B$-band
luminosity  bin  is not  plotted  as  there  is no  secure  morphology
measurement.

%%\clearpage
%%\begin{figure*}
%%\epsscale{0.9}
%%\plotone{UDF_IMG_GOODS_MERGER.eps}
%%\caption{\label{UDF_IMG_GOODS_MERGER} A test of our asymmetry corrections for the
%%GOODS galaxies based on the UDF images. The UDF images are shown for 
%%the GOODS galaxies at $z<$1.0 with probability 
%%for corrected asymmetries to be mergers between 0.3 and 0.7. The uncorrected 
%%asymmetry values indicate non-merger morphologies for all these galaxies while
%%the corrected asymmetry values indicate mergers for half of them, which is
%%consistent with the deep UDF images.}
%%\end{figure*}

%%%%%%%%%%%%%%%%%%%%%%%%%%%%%%%%%%%%%%%%%%%%%%%%%%%%%%%%%%%%%%%%%%%%%%%%%%%%%%%%%%%%%%%%%%%%%%%%%%%%%%%%%%%%%%%%%

\clearpage

\begin{deluxetable}{ccccccccccccccccc}
\tabletypesize{\scriptsize}
\tablecolumns{7}
\tablecaption{\label{Tab_Completeness} The 70\% completeness cut of the morphology-redshift catalog shown in Figure~\ref{Completeness_MOR_RED} }
\tablewidth{0pt}
\tablehead{ \colhead{Redshift}  & \colhead{$R_{\rm gal}$}  & \colhead{$M_{\rm restB}$}  & \colhead{$\mu_{\rm restB}$}    \\
            \colhead{}          & \colhead{[kpc]}           &  \colhead{}               &  \colhead{[mag kpc$^{-2}$]}        }
\startdata
  0.3 & 2.4 & -16.4 & -10.3\\
  0.5 & 3.3 & -17.5 & -10.7\\
  0.7 & 3.9 & -18.6 & -11.9\\
  0.9 & 4.2 & -19.1 & -12.2\\
  1.1 & 4.4 & -19.8 & -12.4\\
\enddata
\tablecomments{The   apparent  galaxy   size,  magnitude   and  surface
brightness  are  labelled  in  Figure~\ref{Completeness_MOR_RED}.  The
absolute magnitude is  measured as $M= m - 5.0$log$(D_{\rm  L}) + 5 +
2.5$log$(1+z)$, where $D_{\rm L}$  is the luminosity distance in pc
and  2.5log(1+z) is the K-correction (see \S~\ref{RED_MOR_RED_CAT}). }
\end{deluxetable}

\begin{deluxetable}{ccccccccccccccccc}
\tabletypesize{\scriptsize}
\tablecolumns{7}
\tablecaption{\label{IRLF_MERGERS} Infrared Luminosity Function of Merger Galaxies }
\tablewidth{0pt}
\tablehead{ \colhead{Log($L_{\rm IR}$ [$h_{70}^{-2}$L$_{\odot}$])}  & \multicolumn{4}{c}{ $\Phi$ [$h_{70}^{3}$Mpc$^{-3}$ LogL$_{\odot}^{-1}$] } \\
            \cline{2-5} \\
            \colhead{   }     &  \colhead{0.4 $<$ $z$ $<$ 0.6} &
            \colhead{0.6 $<$ $z$ $<$ 0.8}  & \colhead{0.8 $<$ $z$ $<$ 1.0}  & \colhead{1.0 $<$ $z$ $<$ 1.2}
}
\startdata 
 
10.75$^{1}$ & ( 4.60$^{+2.27}_{-1.33}$)$\times$10$^{-4}$ &  ...                                      &  ...                                      &  ...                                      \\
11.25$^{1}$ & ( 3.05$^{+1.55}_{-0.96}$)$\times$10$^{-4}$ & ( 3.64$^{+1.77}_{-1.01}$)$\times$10$^{-4}$ & ( 5.66$^{+2.69}_{-1.45}$)$\times$10$^{-4}$ & ( 6.26$^{+3.28}_{-2.13}$)$\times$10$^{-4}$ \\
11.75$^{1}$ & ( 4.23$^{+3.11}_{-2.61}$)$\times$10$^{-5}$ & ( 1.83$^{+0.93}_{-0.58}$)$\times$10$^{-4}$ & ( 4.40$^{+2.10}_{-1.15}$)$\times$10$^{-4}$ & ( 5.58$^{+2.63}_{-1.41}$)$\times$10$^{-4}$ \\
12.25$^{1}$ &  ...                                      & ( 1.56$^{+1.38}_{-1.23}$)$\times$10$^{-5}$ & ( 1.07$^{+0.56}_{-0.37}$)$\times$10$^{-4}$ & ( 1.95$^{+0.96}_{-0.56}$)$\times$10$^{-4}$ \\
12.75$^{1}$ &  ...                                      &  ...                                      &  ...                                      & ( 2.90$^{+1.85}_{-1.44}$)$\times$10$^{-5}$ \\
10.75$^{2}$ & ( 4.60$^{+2.27}_{-1.33}$)$\times$10$^{-4}$ &  ...                                      &  ...                                      &  ...                                      \\
11.25$^{2}$ & ( 3.06$^{+1.55}_{-0.96}$)$\times$10$^{-4}$ & ( 3.54$^{+1.72}_{-0.99}$)$\times$10$^{-4}$ & ( 5.14$^{+2.44}_{-1.33}$)$\times$10$^{-4}$ & ( 5.85$^{+3.11}_{-2.06}$)$\times$10$^{-4}$ \\
11.75$^{2}$ & ( 4.22$^{+3.10}_{-2.61}$)$\times$10$^{-5}$ & ( 1.40$^{+0.73}_{-0.48}$)$\times$10$^{-4}$ & ( 2.74$^{+1.33}_{-0.76}$)$\times$10$^{-4}$ & ( 2.73$^{+1.32}_{-0.75}$)$\times$10$^{-4}$ \\
12.25$^{2}$ &  ...                                      & ( 1.12$^{+1.12}_{-1.03}$)$\times$10$^{-5}$ & ( 7.11$^{+3.94}_{-2.74}$)$\times$10$^{-5}$ & ( 1.11$^{+0.57}_{-0.36}$)$\times$10$^{-4}$ \\
12.75$^{2}$ &  ...                                      &  ...                                      &  ...                                      & ( 1.73$^{+1.27}_{-1.06}$)$\times$10$^{-5}$ \\
 
\enddata 
\tablecomments{ $^{1}$ The galaxy asymmetries are corrected based on the redshifted local LIRGs. 
$^{2}$ The galaxy asymmetries are corrected based on the redshifted low-z ($z$=0.2-0.4) GOODS field galaxies. }
\end{deluxetable}

\begin{deluxetable}{ccccccccccccccccc}
\tabletypesize{\scriptsize}
\tablecolumns{7}
\tablecaption{\label{IRLF_MERGERS_FIT} Parameters of Schetcher Function Fit to IR LFs of Galaxy Mergers}
\tablewidth{0pt}
\tablehead{ \colhead{redshift}  & \colhead{ Log($\Phi_{*}$)}                            & \colhead{ Log($L_{*}$) }  &  \colhead{$\alpha$}   \\
\hline
            \colhead{}          & \colhead{ [$h_{70}^{3}$Mpc$^{-3}$ LogL$_{\odot}^{-1}$]} &  \colhead{ [$h_{70}^{-2}$L$_{\odot}$] }  &  \colhead{}  }
\startdata
0.4$<z^{1}<$0.6 & -3.63$^{+0.06}_{-0.05}$ & 11.37$^{+0.08}_{-0.03}$ & -1.12$^{+0.00}_{-0.00}$ \\
0.6$<z^{1}<$0.8 & -3.70$^{+0.06}_{-0.10}$ & 11.75$^{+0.07}_{-0.00}$ & -1.12$^{+0.00}_{-0.00}$ \\
0.8$<z^{1}<$1.0 & -3.59$^{+0.09}_{-0.03}$ & 12.04$^{+0.17}_{-0.00}$ & -1.12$^{+0.00}_{-0.00}$ \\
1.0$<z^{1}<$1.2 & -3.62$^{+0.02}_{-0.03}$ & 12.29$^{+0.04}_{-0.06}$ & -1.12$^{+0.00}_{-0.00}$ \\
0.4$<z^{2}<$0.6 & -3.63$^{+0.06}_{-0.05}$ & 11.37$^{+0.08}_{-0.04}$ & -1.12$^{+0.00}_{-0.00}$ \\
0.6$<z^{2}<$0.8 & -3.72$^{+0.07}_{-0.11}$ & 11.70$^{+0.07}_{-0.01}$ & -1.12$^{+0.00}_{-0.00}$ \\
0.8$<z^{2}<$1.0 & -3.67$^{+0.04}_{-0.04}$ & 11.98$^{+0.13}_{-0.04}$ & -1.12$^{+0.00}_{-0.00}$ \\
1.0$<z^{2}<$1.2 & -3.80$^{+0.02}_{-0.08}$ & 12.26$^{+0.03}_{-0.07}$ & -1.12$^{+0.00}_{-0.00}$ \\
\enddata 
\tablecomments{ $^{1}$ The galaxy asymmetries are corrected based on the redshifted local LIRGs. 
$^{2}$ The galaxy asymmetries are corrected based on the redshifted low-z ($z$=0.2-0.4) GOODS field galaxies.  }
\end{deluxetable}

\begin{deluxetable}{ccccccccccccccccc}
\tabletypesize{\scriptsize}
\tablecolumns{7}
\tablecaption{\label{MbLF_MERGERS} $B$-band Luminosity Function of Merger Galaxies }
\tablewidth{0pt}
\tablehead{ \colhead{$M_{B}$}  & \multicolumn{5}{c}{ $\Phi$ [$h_{70}^{3}$Mpc$^{-3}$ LogL$_{\odot}^{-1}$] } \\
            \cline{2-6} \\
            \colhead{   }     &  \colhead{0.2 $<$ $z$ $<$ 0.4} &  \colhead{0.4 $<$ $z$ $<$ 0.6} &
            \colhead{0.6 $<$ $z$ $<$ 0.8}  & \colhead{0.8 $<$ $z$ $<$ 1.0}  & \colhead{1.0 $<$ $z$ $<$ 1.2}
}
\startdata

-22.40 &  ...                                       &  ...                                       & ( 6.18$^{+2.34}_{-2.34}$)$\times$10$^{-5}$ & ( 3.09$^{+1.34}_{-1.34}$)$\times$10$^{-5}$ & ( 3.93$^{+1.48}_{-1.48}$)$\times$10$^{-5}$ \\
-21.60 & ( 1.28$^{+0.58}_{-0.58}$)$\times$10$^{-4}$ & ( 9.19$^{+3.53}_{-3.53}$)$\times$10$^{-5}$ & ( 1.84$^{+0.53}_{-0.53}$)$\times$10$^{-4}$ & ( 1.71$^{+0.48}_{-0.48}$)$\times$10$^{-4}$ & ( 2.04$^{+0.54}_{-0.54}$)$\times$10$^{-4}$ \\
-20.80 & ( 1.68$^{+0.69}_{-0.69}$)$\times$10$^{-4}$ & ( 3.31$^{+0.93}_{-0.93}$)$\times$10$^{-4}$ & ( 3.61$^{+0.94}_{-0.94}$)$\times$10$^{-4}$ & ( 4.34$^{+1.09}_{-1.09}$)$\times$10$^{-4}$ & ( 2.98$^{+0.76}_{-0.76}$)$\times$10$^{-4}$ \\
-20.00 & ( 3.48$^{+1.14}_{-1.14}$)$\times$10$^{-4}$ & ( 4.87$^{+1.30}_{-1.30}$)$\times$10$^{-4}$ & ( 4.17$^{+1.07}_{-1.07}$)$\times$10$^{-4}$ & ( 5.74$^{+1.41}_{-1.41}$)$\times$10$^{-4}$ & ( 3.37$^{+0.85}_{-0.85}$)$\times$10$^{-4}$ \\
-19.20 & ( 2.04$^{+0.78}_{-0.78}$)$\times$10$^{-4}$ & ( 3.19$^{+0.90}_{-0.90}$)$\times$10$^{-4}$ & ( 3.42$^{+0.90}_{-0.90}$)$\times$10$^{-4}$ & ( 4.42$^{+1.11}_{-1.11}$)$\times$10$^{-4}$ &  ...                                       \\
-18.40 & ( 2.88$^{+0.99}_{-0.99}$)$\times$10$^{-4}$ & ( 2.48$^{+0.74}_{-0.74}$)$\times$10$^{-4}$ & ( 3.10$^{+0.82}_{-0.82}$)$\times$10$^{-4}$ &  ...                                       &  ...                                       \\
-17.60 & ( 1.93$^{+0.75}_{-0.75}$)$\times$10$^{-4}$ & ( 1.92$^{+0.60}_{-0.60}$)$\times$10$^{-4}$ &  ...                                       &  ...                                       &  ...                                       \\
-16.80 & ( 1.07$^{+0.51}_{-0.51}$)$\times$10$^{-4}$ &  ...                                       &  ...                                       &  ...                                       &  ...                                       \\

\enddata 
%\tablecomments{ }
\end{deluxetable}

\begin{deluxetable}{ccccccccccccccccc}
\tabletypesize{\scriptsize}
\tablecolumns{7}
\tablecaption{\label{MbLF_MERGERS_FIT} Parameters of Schetcher Function Fit to Merger B-band LFs}
\tablewidth{0pt}
\tablehead{\colhead{redshift}  & \colhead{ Log($\Phi_{*}^{0}$)}                  & \colhead{ $M_{B*}^{0}$ }  &  \colhead{$\alpha$}  \\
\hline
           \colhead{}          & \colhead{ [$h_{70}^{3}$Mpc$^{-3}$ M$_{B}^{-1}$]} &  \colhead{ }             & \colhead{}    }
\startdata
0.2$<z<$0.4 &  -3.52$^{+ 0.00}_{- 0.05}$ & -20.72$^{+-0.03}_{--0.13}$ &  -0.53$^{+ 0.00}_{- 0.00}$ \\
0.4$<z<$0.6 &  -3.43$^{+ 0.05}_{- 0.01}$ & -20.63$^{+-0.11}_{--0.01}$ &  -0.53$^{+ 0.00}_{- 0.00}$ \\
0.6$<z<$0.8 &  -3.39$^{+ 0.04}_{- 0.03}$ & -21.05$^{+-0.08}_{--0.00}$ &  -0.53$^{+ 0.00}_{- 0.00}$ \\
0.8$<z<$1.0 &  -3.29$^{+ 0.02}_{- 0.02}$ & -20.77$^{+-0.01}_{--0.06}$ &  -0.53$^{+ 0.00}_{- 0.00}$ \\
1.0$<z<$1.2 &  -3.47$^{+ 0.05}_{- 0.02}$ & -21.04$^{+-0.02}_{--0.03}$ &  -0.53$^{+ 0.00}_{- 0.00}$ \\

\enddata 
%\tablecomments{The B-band luminosity function of merger galaxies are fitted with Schechter Function.  }
\end{deluxetable}


\begin{thebibliography}{}

\bibitem[Abraham et al.(1994)]{Abraham94} Abraham, R. G., Valdes, F., Yee, H. K. C., \& van den Bergh, S. 1994, \apj, 432, 75 

\bibitem[Abraham et al.(1996)]{Abraham96} Abraham, R.~G., van den Bergh, S., Glazebrook, K., Ellis, R.~S., Santiago, B.~X., Surma, P., \& Griffiths, R.~E.\ 1996, \apjs, 107, 1 


\bibitem[Abraham et al.(2003)]{Abraham03} Abraham, R.~G., van den Bergh, S., \& Nair, P.\ 2003, \apj, 588, 218 

\bibitem[Bell et al.(2005)]{Bell05} Bell, E.~F., et al.\ 2005, \apj, 625, 23 

\bibitem[Bershady et al.(2000)]{Bershady00} Bershady, M.~A., 
    Jangren, A., \& Conselice, C.~J.\ 2000, \aj, 119, 2645 

\bibitem[Bertin 
\& Arnouts(1996)]{Bertin96} Bertin, E., \& Arnouts, S.\ 1996, \aaps, 117, 393 

\bibitem[Blanton et al.(2003)]{Blanton03} Blanton, M.~R., et al.\ 2003, \aj, 125, 2348 

\bibitem[Blanton \& Roweis(2007)]{Blanton07} Blanton, M.~R., \& Roweis, S.\ 2007, \aj, 133, 734 

\bibitem[Bridge et al.(2007)]{Bridge07} Bridge, C.~R., et al.\ 2007, \apj, 659, 931 

\bibitem[Brinchmann et al.(1998)]{Brinchmann98} Brinchmann, J., et al.\ 1998, \apj, 499, 112 

\bibitem[Bundy et al.(2004)]{Bundy04} Bundy, K., Fukugita, M., Ellis, R.~S., Kodama, T., \& Conselice, C.~J.\ 2004, \apjl, 601, L123 

\bibitem[Cassata et al.(2005)]{Cassata05} Cassata, P., et al.\ 2005, \mnras, 357, 903 

\bibitem[Cross et al.(2004)]{Cross04} Cross, N.~J.~G., Driver, S.~P., Liske, J., Lemon, D.~J., Peacock, J.~A., Cole, S., Norberg, P., \& Sutherland, W.~J.\ 2004, \mnras, 349, 576 

\bibitem[Conselice et al.(2000)]{Conselice00} Conselice, C.~J., 
  Bershady, M.~A., \& Jangren, A.\ 2000, \apj, 529, 886 

\bibitem[Conselice et al.(2003)]{Conselice03} Conselice, C.~J., 
Bershady, M.~A., Dickinson, M., \& Papovich, C.\ 2003, \aj, 126, 1183 

\bibitem[Conselice et al.(2005)]{Conselice05} Conselice, C.~J., 
Blackburne, J.~A., \& Papovich, C.\ 2005, \apj, 620, 564 

\bibitem[Cowie \& Barger(2008)]{Cowie08} Cowie, L.~L., \& Barger, A.~J.\ 2008, ArXiv e-prints, 806, arXiv:0806.3457 

\bibitem[Daddi et al.(2008)]{Daddi08} Daddi, E., Dannerbauer, H., Elbaz, D., Dickinson, M., Morrison, G., Stern, D., \& Ravindranath, S.\ 2008, \apjl, 673, L21 

\bibitem[De Propris et al.(2007)]{DePropris07} De Propris, R., Conselice, C.~J., Liske, J., Driver, S.~P., Patton, D.~R., Graham, A.~W., \& Allen, P.~D.\ 2007, \apj, 666, 212 

\bibitem[de Ravel et al.(2008)]{deRavel08} de Ravel, L., et al.\ 2008, arXiv:0807.2578 

\bibitem[Faber et al.(2007)]{Faber07} Faber, S.~M., et al.\ 2007, \apj, 665, 265 

\bibitem[Frei et al.(1996)]{Frei96} Frei, Z., Guhathakurta, 
 P., Gunn, J.~E., \& Tyson, J.~A.\ 1996, \aj, 111, 174 

\bibitem[Giavalisco et al.(2004)]{Giavalisco04} Giavalisco, M., et al.\ 2004, \apjl, 600, L93 

\bibitem[Gordon et al.(2005)]{Gordon05} Gordon, K.~D., et al.\ 
2005, \pasp, 117, 503 

%\bibitem[Grazian et al.(2006)]{Grazian06} Grazian, A., et al.\ 2006, \aap, 449, 951

%\bibitem[Hopkins(2004)]{Hopkins04} Hopkins, A.~M.\ 2004, \apj, 615, 209


\bibitem[Kent(1985)]{Kent85} Kent, S.~M.\ 1985, \apjs, 59, 115
 
\bibitem[Le Floc'h et al.(2005)]{LeFloch05} Le Floc'h, E., et al.\ 2005, \apj, 632, 169

\bibitem[Le F{\`e}vre et al.(2000)]{LeFevre00} Le F{\`e}vre, O., et al.\ 2000, \mnras, 311, 565 

%\bibitem[Lilly et al.(1996)]{Lilly96} Lilly, S.~J., Le Fevre, O., Hammer, F., \& Crampton, D.\ 1996, \apjl, 460, L1 

\bibitem[Lin et al.(2008)]{Lin08} Lin, L., et al.\ 2008, \apj, 681, 232 

\bibitem[Lotz et al.(2004)]{Lotz04} Lotz, J.~M., Primack, J., \& Madau, P.\ 2004, \aj, 128, 163 

\bibitem[Lotz et al.(2008a)]{Lotz08a} Lotz, J.~M., et al.\ 2008, ApJ, 672, 177 

\bibitem[Lotz et al.(2008b)]{Lotz08b} Lotz, J.~M., Jonsson, P., Cox, T.~J., \& Primack, J.~R.\ 2008, arXiv:0805.1246 

\bibitem[Melbourne et al.(2008)]{Melbourne08} Melbourne, J., et al.\ 2008, \aj, 135, 1207 

\bibitem[Mihos \& Hernquist(1996)]{Mihos96} Mihos, J.~C., \& Hernquist, L.\ 1996, \apj, 464, 641 

\bibitem[Papovich et al.(2003)]{Papovich03} Papovich, C., Giavalisco, M., Dickinson, M., Conselice, C.~J., \& Ferguson, H.~C.\ 2003, \apj, 598, 827 

\bibitem[Papovich et al.(2004)]{Papovich04} Papovich, C., et al.\ 2004, \apjs, 154, 70 

\bibitem[Patton et al.(2002)]{Patton02} Patton, D.~R., et al.\ 2002, \apj, 565, 208 

\bibitem[P{\'e}rez-Gonz{\'a}lez et al.(2005)]{Perez-Gonzalez05} P{\'e}rez-Gonz{\'a}lez, P.~G., et al.\ 2005, \apj, 630, 82

\bibitem[P{\'e}rez-Gonz{\'a}lez et al.(2008)]{Perez-Gonzalez08} P{\'e}rez-Gonz{\'a}lez, P.~G., et al.\ 2008, \apj, 675, 234 

\bibitem[Petrosian(1976)]{Petrosian76} Petrosian, V.\ 1976, \apjl, 209, L1

\bibitem[Popesso et al.(2008)]{Popesso08} Popesso, P., et al.\ 2008, ArXiv e-prints, 802, arXiv:0802.2930 

\bibitem[Neichel et al.(2008)]{Neichel08} Neichel, B., et al.\ 2008, \aap, 484, 159 

\bibitem[Noeske et al.(2007)]{Noeske07} Noeske, K.~G., et al.\ 2007, \apjl, 660, L47 

%\bibitem[Robertson et al.(2006)]{Robertson06} Robertson, B., Bullock, J.~S., Cox, T.~J., Di Matteo, 
%T., Hernquist, L., Springel, V., \& Yoshida, N.\ 2006, \apj, 645, 986 


\bibitem[Rieke et al.(2009)]{Rieke09} Rieke, G.~H., Alonso-Herrero, A., Weiner, B.~J., P{\'e}rez-Gonz{\'a}lez, P.~G., Blaylock, M., Donley, J.~L., \& Marcillac, D.\ 2009, \apj, 692, 556 

\bibitem[Sanders \& Mirabel(1996)]{Sanders96} Sanders, D.~B., \& Mirabel, I.~F.\ 1996, \araa, 34, 749 

\bibitem[Sanders et al.(2003)]{Sanders03} Sanders, D.~B., 
Mazzarella, J.~M., Kim, D.-C., Surace, J.~A., 
\& Soifer, B.~T.\ 2003, \aj, 126, 1607 

\bibitem[Schmidt(1968)]{Schmidt68} Schmidt, M.\ 1968, \apj, 151, 393 

\bibitem[Shi et al.(2006)]{Shi06} Shi, Y., Rieke, G.~H., Papovich, C., P{\'e}rez-Gonz{\'a}lez, P.~G., \& Le Floc'h, E.\ 2006, \apj, 645, 199 

\bibitem[Shi et al.(2008)]{Shi08} Shi, Y., Rieke, G., Donley, J., Cooper, M., Willmer, C., \& Kirby, E.\ 2008, arXiv:0807.4949 

\bibitem[Springel et al.(2005)]{Springel05} Springel, V., et al.\ 
2005, \nat, 435, 629 

\bibitem[Vanzella et al.(2008)]{Vanzella08} Vanzella, E., et al.\ 2008, \aap, 478, 83 


%\bibitem[Wolf et al.(2003)]{Wolf03} Wolf, C., Meisenheimer, 
%K., Rix, H.-W., Borch, A., Dye, S., \& Kleinheinrich, M.\ 2003, \aap, 401, 73 

\bibitem[Yang et al.(2008)]{Yang08} Yang, Y., et al.\ 2008, \aap, 477, 789 

\bibitem[Zheng et al.(2004)]{Zheng04} Zheng, X.~Z., Hammer, F., Flores, H., Ass{\'e}mat, F., \& Pelat, D.\ 2004, \aap, 421, 847 

\end{thebibliography}
\end{document}